\documentclass[prd,aps,amsfonts,showpacs,superscriptaddress,eqsecnum,nofootinbib,longbibliography,notitlepage]{revtex4-1}

\usepackage{graphicx}
\usepackage[usenames,dvipsnames]{xcolor}
\usepackage{rotating}
\usepackage[tight]{subfigure}
\usepackage[normalem]{ulem}
\usepackage{amsmath,amssymb,graphics,amsthm}
\usepackage{mathtools}

\usepackage[colorlinks=true, urlcolor=violet, linkcolor=blue, citecolor=red, hyperindex=true, linktocpage=true]{hyperref}
\usepackage[capitalise,compress]{cleveref}
\usepackage{soul}

\newcommand{\comment}[1]{}

\makeatletter
\renewcommand{\p@subsection}{}
\renewcommand{\p@subsubsection}{}
\makeatother

\def\CL{\mathcal{L}}

\def\CO{\mathcal{O}}
\def\CP{\mathcal{P}}

\def\CT{\mathcal{T}}

\allowdisplaybreaks

\usepackage{xcolor}
\newcommand{\andy}[1]{{\color{red} [Andy: #1]}}

\usepackage{mathtools}

\newcommand{\ii}{\mathrm{i}}

\newcommand\p{\ensuremath{\partial}}
\newcommand{\be}{\begin{equation}}
\newcommand{\ee}{\end{equation}}
\def\tr{\mathop{\rm tr}}
\newcommand\vep{\varepsilon}
\newcommand{\PG}[1]{\textcolor{magenta}{[Paolo: #1]}}

\newcommand{\ph}{\phantom{b}}

\newcommand{\eqnref}[1]{\eqref{#1}}

\newcommand{\secref}[1]{Section\;\ref{#1}}
\newcommand{\appref}[1]{Appendix\;\ref{#1}}

\begin{document}
\title{Goldstone bosons and fluctuating hydrodynamics with dipole and momentum conservation}

\author{Paolo Glorioso}
\email{paolog@stanford.edu}
\affiliation{Department of Physics, Stanford University, Stanford CA 94305, USA} 

\author{Xiaoyang Huang}
\email{xiaoyang.huang@colorado.edu}
\affiliation{Department of Physics and Center for Theory of Quantum Matter, University of Colorado, Boulder, CO 80309, USA}

\author{Jinkang Guo}
\affiliation{Department of Physics and Center for Theory of Quantum Matter, University of Colorado, Boulder, CO 80309, USA}

\author{Joaquin F. Rodriguez-Nieva}
\affiliation{Department of Physics, Stanford University, Stanford CA 94305, USA}

\author{Andrew Lucas}
\email{andrew.j.lucas@colorado.edu}
\affiliation{Department of Physics and Center for Theory of Quantum Matter, University of Colorado, Boulder, CO 80309, USA}

\begin{abstract}
We develop a Schwinger-Keldysh effective field theory describing the hydrodynamics of a fluid with conserved charge and dipole moments, together with conserved momentum. The resulting hydrodynamic modes are highly unusual, including sound waves with quadratic (magnon-like) dispersion relation and subdiffusive decay rate. Hydrodynamics itself is unstable below four spatial dimensions.  We show that the momentum density is, at leading order, the Goldstone boson for a dipole symmetry which appears spontaneously broken at finite charge density.  Unlike an ordinary fluid, the presence or absence of energy conservation qualitatively changes the decay rates of the hydrodynamic modes.  This effective field theory naturally couples to curved spacetime and background gauge fields; in the flat spacetime limit, we reproduce the ``mixed rank tensor fields" previously coupled to fracton matter.
\end{abstract}

\date{\today}

\maketitle

\tableofcontents

\section{Introduction}
One of the oldest and most successful theories in physics is hydrodynamics.  While hydrodynamics was first understood as a phenomenological set of equations that govern liquids and gases \cite{landau}, over the past century we have instead recognized that hydrodynamics is best understood as the universal effective field theory that governs thermalization in a chaotic many-body system \cite{Crossley:2015evo,Haehl:2015foa,Jensen:2017kzi,Glorioso:2017fpd,liu2018lectures}.  In the simplest scenarios, the degrees of freedom of a hydrodynamic theory correspond to locally conserved quantities;  the way that these modes interact with each other and decay is constrained only by basic symmetries of the theory.  Using modern effective field theory methods, sophisticated nonlinear theories of fluctuating hydrodynamics have been developed and applied to increasingly sophisticated systems.

One family of novel phases of matter which has interesting dynamics arises when the microscopic degrees of freedom are \emph{fractons} -- excitations which are individually immobile, and can only move in tandem \cite{chamon2005quantum,vijay2015new,vijay2016fracton,pretko2017emergent,pretko2017generalized,pretko2017subdimensional,Pretko2018,slagle2017fracton,slagle2018symmetric,schmitz2018recoverable,shirley2019foliated,Seiberg:2020bhn,Distler:2021bop,Gorantla:2020xap,Gorantla:2022eem}.\footnote{While it can be desirable to strengthen this definition to demand that fractons cannot move under the action of any local operator, following the ``fracton hydrodynamics" literature, we will take a looser definition.  (In our theory, a local operator that inserts a quadrupole can move an isolated charge.)}  As a simple example, we can consider a phase of matter in which charge/mass is conserved together with dipole moment/center of mass -- in this case, a single particle cannot move without violating the dipole conservation law!  The past few years have seen an intense study of the fracton phases of matter that can result by combining many of these interacting fractons.  And over the past two years, it has been understood that when such fracton phases thermalize \cite{2020PollmannFragmentation,shattering}, the resulting hydrodynamics is non-trivial \cite{gromovfrachydro,knap2020,morningstar,zhang2020universal}: Fick's law of diffusion, for example, becomes replaced by \emph{subdiffusive} equations, with the dynamical critical exponent dependent on how many multipole moments are conserved.  Many further ``fracton hydrodynamics" universality classes have since been discovered \cite{doshi,iaconis2021,hart2021hidden,sala2021dynamics,knap2021,Burchards:2022lqr,Guo:2022ixk}.

In this paper, we detail a qualitatively new universality class of hydrodynamics that emerges  when fracton-like multipole conservation laws are combined with momentum conservation, which was presented by four of us in a shorter paper \cite{Glorioso:2021bif} --- see also \cite{osborne}. Additionally, we shall discuss the inclusion of energy conservation, also considered in \cite{Grosvenor:2021rrt,glodkowski2022hydrodynamics}, and how it affects dissipative transport.  We focus on the case where dipole moment is the only additional conserved quantity, and where the theory has parity and time-reversal symmetry; in the absence of momentum conservation, the consequences of breaking these symmetries were recently discussed in \cite{Guo:2022ixk}.  Without dipole conservation, such a theory is essentially described by textbook Navier-Stokes equations with incoherent conductivities \cite{Crossley:2015evo}.   With dipole conservation, the Navier-Stokes equations are completely changed \cite{Glorioso:2021bif}.  At finite charge density, the conventional propagating sound modes are replaced by magnon-like propagating modes.   The decay rates of these magnon-like modes is diffusive if energy is conserved, but subdiffusive if energy is not conserved.  And at zero density, the character of the hydrodynamic modes completely changes; the naive derivative expansion of hydrodynamics at finite density is singular as low density is approached. At zero density, assuming particle-hole symmetry, the momentum dynamics decouples from that of charge within linear response. The character of collective modes in this case is completely different, where momentum and charge display diffusive and subdiffusive damping, respectively.

The subtle nature of this emergent hydrodynamics is intricately related to the fact that (in quantum mechanics) the dipole moment operator $D$, and net momentum operator $P$, do not commute \cite{gromov2018towards}: \begin{equation}
    [D,P] = \mathrm{i}Q, \label{eq:DPQ}
\end{equation}
where $Q$ represents total charge.  An analogous classical statement holds for Poisson brackets. One might expect that such a commutation relation is similar to angular momentum commutation relations in an isotropic fluid -- such commutation relations lead not to new propagating degrees of freedom, but rather constraints on the currents of other modes (the stress tensor, in this case).  However, at finite density, (\ref{eq:DPQ}) implies that momentum susceptibility (the generalization of mass density) is singular!  This means that a naive hydrodynamic degree of freedom -- fluid velocity -- is non-local.  One of the main results of this paper is that we can nevertheless construct a \emph{local} hydrodynamic theory, using unconventional degrees of freedom.  In Section \ref{sec:EFT}, we will describe how to construct this EFT following the constructions of \cite{Crossley:2015evo,Haehl:2015foa,Jensen:2017kzi,Glorioso:2017fpd,liu2018lectures}.  In the process, we comment on the coupling of this theory to background geometry (vielbein), although much of this technical work is relegated to appendices.  Following \cite{slaglecurve,Jain2021,Bidussi:2021nmp,Pena-Benitez:2021ipo}, we hope this can further stimulate work on understanding how and when fractons can be coupled to gravity.

As reported in \cite{Glorioso:2021bif}, these hydrodynamic theories can be \emph{unstable below four dimensions}. This is true both without energy conservation, and with energy conservation at infinite temperature (under mild assumptions). This result generalizes the well-known Kardar-Parisi-Zhang instability of an equilibrium fluid (without dipole conservation) in one dimension \cite{KPZ}, and implies the existence of a non-equilibrium fixed point in three dimensions, in an undriven system.   

From many perspectives, we will show that the commutation relation (\ref{eq:DPQ}) implies there is \emph{spontaneous symmetry breaking}.  In a finite density state, $D$ and $P$ do not commute, so they cannot be diagonalized simultaneously -- in a state with fixed momentum, there are large fluctuations in dipole moment.  Unlike more conventional \emph{compact} non-Abelian symmetry groups such as SU(2), here one cannot find \emph{any} physical ``singlet" states in the Hilbert space which lie in trivial representations.   (This is analogous to textbook quantum mechanics: one cannot find simultaneous eigenstates of $x$ and $p$.)  So it seems that trivially, dipole and/or momentum will be spontaneously broken, in agreement with previous literature on low-dimensional SSB with non-compact symmetry groups \cite{Niedermaier:2003cq}. In this paper, we focus on ensembles with fixed momentum density (which seems more physical to us).  A consequence is that the propagating momentum density is (at leading order in the hydrodynamic expansion) proportional to the \emph{Goldstone boson} for  broken dipole symmetry.  
In one and two space dimensions, the fluctuations of this Goldstone boson are very large, and for this reason a recent work \cite{Kapustin:2022fzp} proved that there is no SSB within the context of the Mermin-Wagner theorem. On the other hand, we will see that the hydrodynamics in low dimension a single hydrodynamic mode still contains all of the spectral weight required to saturate the Goldstone theorem. The presence of this Goldstone boson in the hydrodynamic theory suggests an unusual paradigm for possible ``spontaneous symmetry breaking.''
Following recent discussions on the spontaneous breaking of boost symmetry in fluids \cite{nicolisSSB, komargodski_boost}, we will discuss at some length the nature of the apparent spontaneous symmetry breaking in Section \ref{sec:ssb}.


In Section \ref{sec:energy}, we extend the discussion of the EFT to models with energy conservation.  The observation of interest is that energy conservation \emph{changes the dynamical universality class} of the dipole-momentum conserving fixed point.\footnote{Interestingly, there was a debate in past literature about the role of energy conservation in flowing from the Navier-Stokes to the KPZ fixed point \cite{Lepri_heat_1997,Lee_heat_2010,Zhong_heat_2012}. The consensus is now that energy conservation indeed does not disturb the KPZ point \cite{Das_heat_2014}.  What was missing in the past was just to incorporate dipole conservation!} Intuitively, this can be understood as follows:  energy can diffuse, while charge must subdiffuse.   Due to thermodynamics at finite charge and energy density, however, charge and energy modes generically ``mix" (e.g. the propagating sound wave would involve both charge and energy fluctuations).   As a consequence, the dominant decay channel is always through energy diffusion, which leads to $z=2$, rather than $z=4$, at the hydrodynamic (Gaussian) fixed point.

Lastly, in Section \ref{sec:gal}, we discuss how a dipole-conserving theory can arise in the infinite mass limit of a theory with Galilean invariance.  This may suggest one way to look for this physics in experimental systems, e.g. in (nearly) flat bands \cite{Jo_Kagome,Mukherjee_photonic,taie_2015_coherent,cao_2018_unconventional,cao_2020_strange} in condensed matter systems.   

Several complementary discussions and technical computations are included in the appendices. In \appref{sec:MM}, the memory matrix formalism is applied to derive the normal modes and diverging momentum susceptibility. \appref{app:curved} consists of a detailed derivation of dipole-conserving hydrodynamics in curved spacetime. In \appref{app:consistency}, the consistency between dipole symmetry and geometry is verified. 

\emph{Note Added:} After this paper was released, we became aware of 
another recent work \cite{Glodkowski:2022xje}, which contains complementary results to our Section \ref{sec:energy}.

\comment{

\section{Landau route to hydrodynamics}\label{sec:landau}
In this section, we use the canonical arguments based on the second law of thermodynamics to derive the hydrodynamics of conserved momentum, charge, and dipole. The fundamental assumption of hydrodynamics is that the late time physics is governed by the conserved quantities of the system, which we shall write as
\be P^i = \int d^d x\, \pi^i,\quad Q=\int d^d x\, n \ \ee
where $\pi^i$ and $n$ are the momentum and charge density, respectively. 

We have not included the dipole density as a separate degree of freedom, as the dipole charge is determined by the charge density itself: $D^i=\int d^d x\, x^i n$. [EXPLAIN WHY] 

$n$ and $\pi^i$ are subject to local conservation laws:
\be \label{eom1}\p_t \pi^i + \p_j T^{ji}=0,\qquad \p_t n+\p_i J^i=0\ ,\ee
where $T^{ij}$ and $J^i$ are stress and charge flux, and are assumed to be local in $\pi^i,n$. Crucially, we also need to demand
\be\label{dipfl} J^i = \p_j J^{ji}\ ,\ee
which comes from dipole conservation:
\be \p_t \int d^d x x^i n= -\int d^d x x^i\p_j J^j=\int d^d x J^i\ ,\ee
where the right-hand side vanishes only if $J^i$ satisfies (\ref{dipfl}). We will also demand that $J^{ij}$ be local in the densities. Eqs. (\ref{eom1}),(\ref{dipfl}) completely specifies the time evolution of $\pi^i$ and $n$.

\subsection{Ideal hydrodynamics}

We now impose that the dynamics of these densities be consistent with the local second law of thermodynamics. This amounts at finding a vector $S^\mu$ such that $\p_\mu S^\mu\geq 0$ when evaluated on solutions to hydrodynamics, where the time component $S^t$ coincides with the thermodynamic entropy density, $S^t=s$, if fields are restricted to be homogeneous. This basic constraint will uniquely determine the concrete expressions of $T^{ij},J^i$ in terms of $\pi^i,n$, order by order in derivatives, up to phenomenological coefficients that are determined by the specific underlying system. We shall implement this procedure order by order in derivatives. Contrary to all cases known to us, we will see that this hydrodynamics is special in that, first, the homogeneous part of momentum density decouples from the dynamics, and second, there is an emergent scale determined by the background charge density of the system.

We begin by first assuming that the entropy density is a function of momentum and charge densities $s=s(\pi^i,n)$, and we will see that this leads to breaking relation (\ref{dipfl}). Recall the thermodynamic relation $Tds= -V^i d\pi^i-\mu dn$, where $V^i$ and $\mu$ are the velocity and chemical potential of the system. The temperature $T$ is set to be a constant, since we are neglecting energy conservation. The generalization of this relation to ideal hydrodynamics reads
\be \label{2ndla0}T\p_\mu S^\mu =-V^i\p_\mu T^{\mu i} - \mu\p_\alpha J^\alpha. \ee
The most general expressions for the fluxes are $S^i = s_1 V^i$, $T^{ij} = p \delta^{ij} + h_1 V^i V^j$, $J^i = h_2 V^i$.
Plugging these expressions in (\ref{2ndla0}) gives $s_1=s$, $p=Ts+\mu n+ V^i \pi^i$, $h_1 V^i =\frac{\p s}{\p V^i}=\pi^i$ and $h_2=n$, which are the current constitutive relations of standard charged hydrodynamics, and indeed we have not used anywhere the fact that we are dealing with a dipole-conserving fluid. We see that  (\ref{dipfl}) implies the relation $n V^i = \p_j J^{ji}$, which expresses that $J^{ij}$ is non-local in the hydrodynamic variables, against our assumptions. 

The only way for (\ref{dipfl}) to be consistent with locality, is to demand 
\be s=s(\p_i v_j,n),\qquad v_i=\frac{\pi^i}{n}\ .\ee
Repeating the analysis above, this time taking
\be S^i = s_1 V^i+\Delta S^i,\quad  T^{ij} = p \delta^{ij} + h_1 V^i V^j+\Delta T^{ij},\quad J^i = h_2 V^i +\Delta J^i\ ,\ee
where $\Delta S^i,\Delta T^{ij},\Delta J^i$ are higher-derivative expressions of $v_i$ and $n$ to be determined. Plugging in (\ref{2ndla0}), gives
\begin{gather} T^{ij}=p \delta^{ij}+ V^i \pi^j-\psi_{ik}\p_j v_k,\qquad J^i= n V^i\\
\beta^i=\frac 1n \p_j \psi_{ji},\qquad \mu=-\p_n s-V^i v^i\\
p=s-n\p_n s,\qquad S^i=n\p_n sV^i-\psi_{ij}\p_t v_j \ \text{ \PG{ check this!}}\ ,
\end{gather}
where we defined $\psi_{ij}=\left.\frac{\p s}{\p(\p_i v_j)}\right|_n$. 

Note that, the entropy density $s$ as well as equations (\ref{eom1}) are invariant under the shift 
\be\label{dipsh} \pi^i\to \pi^i + n c^i,\qquad T^{ij}\to T^{ij} + J^j c^i\ ,\ee where $c^i$ is a constant vector. This invariance is a manifestation of the dipole algebra (\ref{eq:DPQ}). Indeed, eq. (\ref{eq:DPQ}) implies, using locality: $[D,\pi^i]=i n$, thus leading to (\ref{dipsh}). 

\subsection{Dissipative hydrodynamics}
[dissipative hydro]

\subsection{The charge neutral limit}

Note that the equation determining $\beta^i=\frac 1n \p_j \psi_{ji}$ appears to become singular as we approach charge neutrality $n\to 0$. To be more concrete, let us expand in linear perturbations around finite density, $n=n_0+\delta n$, and treat $\delta n,\pi^i,\beta^i$ as infinitesimal. Then $s=\frac 1{2n0^2} a_4^{ijkl}\p_i \pi_j \p_k \pi_l+\cdots$, where the dots denote terms depending on $\delta n$, and $\beta^i=\frac 1{n_0^2} a_4^{jikl}\p_j\p_k \pi_l$. As $n_0\to 0$, and assuming the underlying system is charge-conjugation invariant, we expect charge and momentum dynamics to decouple, where in particular the momentum dynamics, being insensitive to the presence of charge conservation, is expected to be that of a standard fluid: we expect that momentum will display diffusive behavior. This intuition can be reconciled with our results above by noticing a subtle change in the derivative expansion at charge neutrality. To see this, we simply need to inspect higher-derivative terms: $s=\frac 1{2n0^2} a_4(\p_i \pi_j)^2+\frac{a_5}{n_0^4} (\p_j^2 \pi_i)^2+\cdots$, where we are neglecting possible tensor structures $a_4^{ijkl},a_5^{ijklpq}$ for simplicity, and where the factor of $n_0^{-4}$ multiplying $a_5$ will be justified below. Repeating the analysis around (\ref{2ndla0}) one finds $\beta^i = \frac 1{n_0^2} a_4 \p^2_j \pi_i-\frac {a_5}{n_0^4}\p_j^4 \pi_i$. Solving for momentum, we get $\p_j^2\pi_i=\frac {1}{a_4}\left(n_0^2\beta^i+\frac{a_5 }{a_4}\p_j^2\beta^i\right)$. Now, as $n_0\to 0$, keeping $\pi^i$ fixed we then see that $\pi^i\to \frac {a_5}{a_4^2}\beta^i$, precisely leading to decoupled momentum conservation with diffusive behavior, according to [dissipative eq.].

Note that, had we chosen a different scaling for $a_5$, say $a_5\sim n_0^2$, the charge neutrality limit would require a divergent $\beta^i$, leading to a singular limit for the momentum conservation equation. We will see how these conclusions are reached in a straightforward way using the effective field theory approach of sec. \ref{sec:EFT}.
}

\section{Effective field theory of hydrodynamics} \label{sec:EFT}
One main result is that hydrodynamics with dipole conservation possesses anomalous scaling, which is due to the interplay between the nonlinear hydrodynamic interactions and hydrodynamic fluctuations \cite{Glorioso:2021bif}. To derive this we shall use a recently formulated effective field theory (EFT) of hydrodynamics, which systematically describes fluctuations by encoding hydrodynamics into an effective action \cite{Crossley:2015evo,Haehl:2015foa,Jensen:2017kzi,Glorioso:2017fpd,liu2018lectures}. 

\subsection{General setup}
The aim of the EFT approach is to systematically encode the correlation functions of hydrodynamic densities and currents. Such correlation functions have the general form
\be \label{pathin}\mathrm{Tr}(\mathcal T(J_1J_2\cdots)\rho_0\tilde {\mathcal T}(J_3 J_4\cdots\cdots))=\int_{\rho_0} D\psi_1 D\psi_2 \,e^{iS_0[\psi_1]-iS_0[\psi_2]}\,J_1[\psi_1]J_2[\psi_1]J_3[\psi_2]J_4[\psi_2]\cdots\ ,\ee
where, in the first expression, $\mathcal T$ and $\tilde{\mathcal T}$ denote time- and anti time-ordering, $\rho_0$ is the initial state, which we take to be thermal $\rho_0=e^{-\beta H}/\tr(e^{-\beta H})$, with $H$ the microscopic Hamiltonian of the system, and $J_1,J_2,\dots$ are operators inserted at $(t_1,\vec x_1),(t_2,\vec x_2),\dots$. On the right-hand side, we formally rewrote the correlator as a path-integral, where $S_0$ is the action of the microscopic dynamics, and $\psi_1,\psi_2$ are a doubled copy of the degrees of freedom of the system. Since on the left-hand side we have a forward (backward) time evolution given by the time-ordered (anti-time ordered) product, the path integral contains two exponentials of the action $S_0$, with a relative minus sign, as the first one corresponds to forward evolution, while the second one to backward evolution. In other words, the doubling of degrees of freedom comes from that the evolution of the density matrix $\rho_0 \to U(t)\rho_0 U^\dag(t)$ contains two factors of the evolution, one forward and one backward. Computing hydrodynamic correlation functions from the microscopic dynamics is very hard. We thus want to introduce an EFT approach that substitutes the right-hand side of (\ref{pathin}) with a simpler action:
\be \label{pathin1}\mathrm{Tr}(\mathcal T(J_1J_2\cdots)\rho_0\tilde {\mathcal T}(J_3 J_4\cdots\cdots))=\int D\chi_1 D\chi_2 \,e^{iS[\chi_1,\chi_2]}\,J_1[\chi_1]J_2[\chi_1]J_3[\chi_2]J_4[\chi_2]\cdots\ ,\ee
where $S$ is the effective action for hydrodynamics, and $\chi_1,\chi_2$ denote the doubled hydrodynamic degrees of freedom. The action $S$ will encode the effects of fluctuation and dissipation and, in particular, will allow us to predict the existence of anomalous scaling.

We shall now introduce the degrees of freedom of this EFT. These should be fields that nonlinearly realize the symmetries associated to conservation of charge, dipole and momentum. For momentum conservation, we introduce a set of coordinate fields $X^i=X^i(\sigma^t,\sigma^I)$ which nonlinearly realize translations $P^i$, i.e.
\be \label{tr1} X^i(\sigma^t,\sigma^I)\to X^i(\sigma^t,\sigma^I)+\xi^i\ ,\ee
where $\xi^i$ is a constant vector. We are using $\sigma^I$ to denote an auxiliary coordinate system which can be thought of as labeling the fluid parcels at a fixed value of time $\sigma^t$.\footnote{In older literature, these are the so-called ``Lagrangian specification'' of the fluid \cite{Nicolis:2013lma}.} The coordinates $X^i(\sigma^t,\sigma^I)$ describe the trajectory of the fluid parcel labeled by $\sigma^I$ as a function of time $\sigma^t$. The coordinates $(\sigma^t,X^i)$ are the ``physical'' ones, in the sense that they label the time and space in the lab reference frame.\footnote{It is convenient to denote time by $\sigma^t$ as, in what follows, we will often need to take derivatives with respect to time at fixed $\sigma^I$, not at fixed $X^i$.}

Next, we also have a vector degree of freedom $\varphi^i(\sigma^t,\sigma^I)$ that nonlinearly realizes the dipole shift symmetry $D^i$:
\be \label{tr2} \varphi^i(\sigma^t,\sigma^I)\to \varphi^i(\sigma^t,\sigma^I)+c^i\ ,\ee
where $c^i$ is a constant vector. Finally, for charge $Q$, the associated degree of freedom is a scalar $\varphi(\sigma^t,\sigma^I)$, and transforms as
\be \label{tr3} \varphi(\sigma^t,\sigma^I)\to \varphi(\sigma^t,\sigma^I)+a-c^i X^i\ ,\ee
where $a$ is a constant denoting the parameter of transformations associated to $Q$. The fields $\varphi$ and $\varphi^i$ can be heuristically viewed as describing the ``local phase'' of the fluid $e^{i(\varphi+X^i\varphi^i)}$, where this particular form is motivated from the fact that, for dipole-conserving field theories, $U(1)$ global transformations can have a linear dependence in spatial coordinates \cite{pretko2017generalized}. Note that $\varphi$ transforms also under dipole shifts. This particular transformation rule is implied by the commutator (\ref{eq:DPQ}). Indeed, writing infinitesimal translation and dipole shift as $\delta_\xi \varphi=\xi^i\p_i\varphi$, $\delta_{c} \varphi=-c^iX^i$, we have
\be (\delta_c\delta_\xi-\delta_\xi\delta_c) \varphi=c^i\xi^i\ ,\ee
i.e. the commutator is an infinitesimal shift of $\varphi$, as required by (\ref{eq:DPQ}). It can also be verified that the last term in (\ref{tr3}) is the most general transormation consistent with (\ref{eq:DPQ}). The effective action will be invariant under transformations (\ref{tr1}), (\ref{tr2}), and (\ref{tr3}) which, as a consequence of Noether's theorem, correspond to the statement of conservation of momentum, dipole and charge, respectively. Moreover, we provide a thorough and careful consistency check of the symmetry algebra in \appref{app:consistency}.

Now recall from above that all the degrees of freedom have to be doubled, so we will have $X_1^i,X_2^i,\varphi_1^i,\varphi_2^i,\varphi_1$ and $\varphi_2$. The symmetries (\ref{tr1}), (\ref{tr2}), and (\ref{tr3}) will also be doubled, which in turn correspond to the conservation of the corresponding hydrodynamic currents defined in the forward and backward time contours. Unlike in the path integral (\ref{pathin}), the effective action appearing in (\ref{pathin1}) does not have a factorized form. This is because, as a result of the coarse-graining, where the fast-moving degrees of freedom have been integrated out, new couplings that are local in the ``folded'' time have been generated. These cross-couplings are responsible for dissipations and fluctuations. While the effective action loses factorization, it still satisfies several properties that come from the unitarity of the underlying microscopic evolution \cite{Crossley:2015evo,Glorioso:2017fpd}: 
\be S[\chi,\chi]=0,\qquad S[\chi_2,\chi_1]=-S^*[\chi_1,\chi_2],\qquad \text{Im}\,S[\chi_1,\chi_2]\geq 0\ ,\ee
where $\chi_1,\chi_2$ collectively denote the two copies of $X^i,\varphi^i,\varphi$. Note in particular that the action can (and will) be complex-valued; as we will see this is a basic consequence of having thermal fluctuations. Additionally, since the initial state $\rho_0$ is thermal, and assuming that the microscopic Hamiltonian $H$ is invariant under time-reversal, the effective action satisfies a discrete $\mathbb Z_2$ symmetry called ``dynamical KMS symmetry'': 
\be\label{kms} S[\chi_1,\chi_2]=S[\tilde\chi_1,\tilde\chi_2],\qquad \tilde \chi_1(\sigma^t,\sigma^I)=(-1)^\eta \chi_1(-\sigma^t,\sigma^I),\quad \tilde \chi_2(\sigma^t,\sigma)=(-1)^\eta \chi_2(-\sigma^t-i\beta,\sigma^I)\ ,\ee
where $(-1)^\eta=\pm 1$ denotes the time-reversal eigenvalue of $\chi$.
This symmetry is equivalent to the Euclidean time periodicity of correlation functions on a thermal state with inverse temperature $\beta$. In our effective action, it will relate couplings responsible for dissipation with those describing fluctuations, and it will ensure consistency with the second law of thermodynamics, Onsager relations, and existence of equilibrium. Eq. (\ref{kms}) can be extended to situations where the microscopic Hamiltonian is invariant under a more general discrete symmetry, so long as such symmetry contains time-reversal. A proof of (\ref{kms}) is given in \cite{Crossley:2015evo}.

To complete our effective field theory, we need an additional set of symmetries that characterize the fact that the late-time behavior of the system is that of a fluid. Recall that $\sigma^I$ should be interpreted as labels of fluid elements at a fixed value of $\sigma^t$. Adiabatically reshuffling fluid elements has a vanishing cost in energy, since, in contrast to a solid, fluid parcels are not pinned to a particular spatial location. This means that a specific way to label fluid elements at a given time is not physical, and thus the effective action should be invariant under time-independent redefinitions of $\sigma^I$:
\be \label{eq:relabelspace}
\sigma^I\to \sigma^{'I}(\sigma^J)\ .\ee
Had we not considered this symmetry, the action could depend on arbitrary derivatives $\p_IX^i$, and we would describe a solid instead of a liquid. Analogously, in the charge sector, we have the freedom to relabel the local phase $e^{i(\varphi+X^i\varphi^i)}$ at a fixed time. This amounts to requiring the symmetry
\be\label{diags} \varphi_1(\sigma^t,\sigma^I)\to \varphi_1(\sigma^t,\sigma^I)+\lambda(\sigma^I),\qquad \varphi_2(\sigma^t,\sigma^I)\to \varphi_2(\sigma^t,\sigma^I)+\lambda(\sigma^I)\ ,\ee
where $\lambda(\sigma^I)$ is a time-independent redefinition of the phase and can be arbitrarily assigned on each fluid element $\sigma^I$. The symmetry (\ref{diags}), dubbed \emph{diagonal shift symmetry} in \cite{Crossley:2015evo,Glorioso:2017fpd}, states the absence of spontaneous symmetry breaking of the global $U(1)$ symmetry. Indeed, in the occurrence of spontaneous symmetry breaking, the full information about the phase would be a physical (of course, up to constant shifts of the phase), which would give rise to a superfluid. Instead, in the present context, we are merely interested in the conservation of charge (and dipole) in the absence of spontaneous symmetry breaking of charge.

Unlike $\varphi_{1,2}$, $\varphi^i_{1,2}$ does not have a diagonal shift symmetry. Indeed, one can show that imposing the symmetry $\varphi^i_{1,2}\to \varphi^i_{1,2} + \lambda^i(\sigma^I)$ leads to the incorrect Ward identities, as we will show below (\ref{eq:WIdipolemain}) in the following Section.  We will later see in Sec. \ref{sec:ssb} that this can be understood as a consequence of spontaneous symmetry breaking --- $\varphi^i$ is a Goldstone boson. 

\subsection{Classical limit and hydrodynamic effective theory}

The formalism we have introduced above is based on quantum mechanics. In the present paper, however, we are interested in the emergent classical, high-temperature hydrodynamic behavior of many-body systems. There is a simple way to take the classical limit of this framework which retains the physics we are interested in and has the benefit of considerably simplifying various technical aspects. To this aim, we restore factors of $\hbar$ and write $\chi_1=\chi+\frac 12 \hbar \chi_a$, $\chi_2=\chi-\frac 12 \hbar \chi_a$, where aggain $\chi$ collectively denotes the hydrodynamic fields, for example: $X_1^i=X^i+\frac 12 \hbar X_a^i$, etc. The fact that $\chi_1-\chi_2$ is linear in $\hbar$ can be heuristically understood from the fact that the forward and backward time evolutions are located a distance $\hbar\beta$ from each other, and thus, as $\hbar\to 0$, $\chi_1-\chi_2$ should vanish linearly in $\hbar$. In this limit, the dynamical KMS symmetry becomes
\be \tilde \chi(\sigma^t,\sigma^I)=(-1)^\eta\chi(-\sigma^t,\sigma^I),\qquad 
\tilde \chi_a(\sigma^t,\sigma^I)=(-1)^\eta\{\chi_a(-\sigma^t,\sigma^I)+i\beta \p_t\chi(-\sigma^t,\sigma^I)\}\ ,\ee
where the dependence on $\hbar$ has factorized out, and the nonlocal time shift in (\ref{kms}) reduced to an \emph{exact} time derivative, allowing for a more straightforward implementation.

We now proceed to writing down the invariant blocks that will be used to write the effective action. We assume rotational invariance, but a generalization to discrete rotational symmetry is straightforward \cite{Huang:2022ixj}. We recall that the energy conservation is not assumed, so we take $\beta_0$ as a constant inverse temperature.\footnote{The recent formalism of \cite{Guo:2022ixk}, which can build effective field theories for non-thermal systems (i.e. those whose steady state is not of the form $\exp[-\beta H]$, may allow us to put this construction on a firmer footing.  However, it is not known how to incorporate the non-Abelian multipole algebra or spontaneous symmetry breaking into this formalism.  Revisiting this question would be interesting in future work.} We will come back to include energy conservation in \secref{sec:energy}. For completeness, we summarize the notation here; see \appref{app:curved} for more details. We denote $\mu,\nu=0,1,2,3$ for the physical spacetime, and $A,B=t,x,y,z$ for the fluid spacetime, and use $i,j$, $I,J$ to indicate their spatial subspace, respectively. We introduce the internal spacetime indices $\alpha,\beta$ and $b,c$ for its spatial subspace (we reserve $a$ to describe $a$-fields in the Keldysh contour!).

To incorporate the gauge invariance, we introduce the background gauge field $e^b_\mu$, $A_\mu$ and $A^b_{\mu}$, such that the invariant building-blocks in the fluid spacetime are defined as ($s=1,2$)
\begin{subequations}\label{eq:block}
\begin{align}
    e^b_{s,A}(\sigma) &= \frac{\p X^\mu_s(\sigma)}{\p \sigma^A} e^b_{s,\mu}(\sigma), \\
    B_{s,A}(\sigma) &= \frac{\p X^\mu_s(\sigma)}{\p \sigma^A}\left( A_{s,\mu}(\sigma)+e^b_{s,\mu}(\sigma)\varphi_{s,b}(\sigma)\right)+\frac{\p \varphi_s(\sigma)}{\p \sigma^A},\\
    K_{s,A}^b(\sigma)&=  \frac{\p X^\mu_{s}(\sigma)}{\p \sigma^A}\left(A_{s,\mu}^b+\omega_{s,\mu c}^b\varphi_s^c \right)+\frac{\p \varphi^b_s(\sigma)}{\p \sigma^A}.
\end{align}
\end{subequations}
See a derivation in \appref{app:curved}. 
In the main text, we will be particularly interested in the geometry where $e^b_{1\mu}+e^b_{2\mu} = 2\delta^b_\mu$ and $\omega^b_{s,\mu c}=0$. In the classical limit, this corresponds to working with a flat spacetime and allowing $e^b_{a,\mu} = e^b_{1\mu}-e^b_{2\mu}$ to source the stress tensor.
Throughout this section, we take $e^0_{s\mu}=\delta^0_{s\mu}$, but will consider a more general background when evaluating the energy fluctuations in \secref{sec:energy}. 
We denote the $r,a$-fields as follows
\begin{subequations}
\begin{align}
    \Lambda_{r} &= \frac{\Lambda_1+\Lambda_2}{2},\\
    \Lambda_{a} &=\Lambda_1-\Lambda_2,
\end{align}
\end{subequations}
where $\Lambda_{r,a}$ denote collectively the background and dynamical fields. The $r$-fields of the blocks are
\begin{subequations}\label{eq:blockr}
\begin{align}
    e^b_{r,A} &= \p_A X^\mu e^b_\mu, \\
    B_{r,A} &= \p_A \varphi+\p_A X^\mu A_\mu  + \p_A X^\mu e^b_\mu \varphi_b, \\
    K^b_{r,A} &= \p_AX^\mu K^b_{r,\mu} = \p_AX^\mu(\p_\mu \varphi^b+A^b_\mu) ,\label{eq:KK}
\end{align}
\end{subequations}
While $a$-fields are always invariant under the relabeling symmetries \eqnref{eq:relabelspace} and \eqnref{diags}, the $r$-fields are not, and the invariant $r$-fields without derivatives are
\begin{align}
    e^b_{r,t} = \p_t X^\mu e^b_\mu\equiv \beta_0 u^\mu e^b_\mu =\beta_0 u^b,\quad
    B_{r,t} = \beta_0 u^\mu B_{r,\mu} \equiv  \beta_0 \mu,\quad K^b_{r,\mu},
\end{align}
from which we defined the thermodynamic variables $u^\mu,\mu$ and $K^b_{r,\mu}$. 
Below, we omit the index $r$ for simplicity.
In the classical limit and physical spacetime, the $a$-fields of the invariant blocks can be written as 
\begin{subequations}
\begin{alignat}{3}
    e^b_{a,A}&= \p_A X^\mu E^b_{a,\mu},\quad  && E^b_{a,\mu}= e^b_{a,\mu}+\CL_{X_a}e^b_\mu, \\
    B_{a,A}&= \p_A X^\mu C_{a,\mu},   &&C_{a,\mu} = A_{a,\mu}+\p_\mu \varphi_a+\CL_{X_a}A_\mu+e^b_\mu \varphi_{a,b}+E^b_{a,\mu}\varphi_b, \\
    K^b_{a,A} &= \p_A X^\mu K^b_{a,\mu},  && K^b_{a,\mu}=\p_\mu \varphi^b_a + A^b_{a,\mu}+\CL_{X_a}A^b_\mu.
\end{alignat}
\end{subequations}

To the leading order in $a$-fields, the effective Lagrangian is given by
\begin{align}\label{eq:linearL}
    \CL = \hat{T}^\mu_b E^b_{a,\mu} +J^\mu C_{a,\mu}+J^{\mu}_b K^b_{a,\mu}+\ldots.
\end{align}
Note that in this case the stress tensor is not equal to the coefficient $T^\mu_b \neq \hat{T}^\mu_b$.  Indeed, the Ward identities in the absence of stochastic fluctuations are obtained by varying $\CL$ with respect to $X^i_a$, $\varphi_a$ and $\varphi^b_{a}$ and then setting $a$-fields to zero. This leads to
\begin{subequations}\label{eq:WImain}
\begin{align}
    \p_\mu (\hat{T}^\mu_b+J^\mu \varphi_b) e^b_i 
    +A^b_i e_{\mu b} J^\mu -F^b_{i \mu} J^\mu_b - F_{i\mu}J^\mu &=0,\label{eq:WImommain} \\
    \p_\mu J^\mu  &=0,\\
    \p_\mu J^{\mu}_b  - J^\mu e_{\mu b} &=0,\label{eq:WIdipolemain}
\end{align}
\end{subequations}
where $F_{\mu\nu}=\p_\mu A_\nu - \p_\nu A_\mu$ and $F^b_{\mu\nu} = \p_\mu A^b_\nu - \p_\nu A^b_\mu$ are the $U(1)$ and dipole field strength, respectively.
In the following, we will construct the effective field theory to determine the stress tensor and currents. We will see that the stress tensor is indeed given by the term in the bracket in \eqnref{eq:WImommain}.

Note that, had we imposed diagonal shift symmetry on the dipole field $\varphi_b\to\varphi_b+\lambda_b(\sigma^I)$, this would affect the structure of the Ward identities. Indeed, neglecting background fields, $C_{a,\mu}=\p_\mu \varphi_a+e^b_\mu\varphi_{a,b}+\p_\mu X^b_a\varphi_b$, and we see that $C_{a,\mu}$ would transform nontrivially under $\lambda_b$. This in turn would affect the structure of the action (\ref{eq:linearL}) and thus alter the Ward identities. Ward identities should only be determined in terms of the global symmetries of the system (or their gauged version), i.e. eqs. (\ref{tr1})-(\ref{tr3}), therefore it is \emph{necessary} that $\varphi_b$ does not possess a diagonal shift symmetry, unlike $\varphi$.

Additionally, we note that, unlike most studies about dipole field theory (e.g. \cite{jensencurved}), the dipole gauge field $A^b_\mu$ as well as the dipole current $J^\mu_b$ by no means need to be symmetric in their spatial indices. 
In dipole hydrodynamics \emph{without} momentum, $J_{ij}\sim\p_i\p_j \mu$, which automatically decouples the antisymmetric part at this derivative order. The momentum density, on the other hand, can contribute to the antisymmetric part of $J_{ij}$. As we will see later, such antisymmetric term contributes to the momentum subdiffusion mode in \eqnref{eq:normalmom} through the coefficient $a_3$ that is defined in \eqnref{eq:atensor}, and thus has physical consequences.

At last, let us consider a system that preserves the symmetry $\Theta = \CP\CT$, where $\CP$ acts as flipping all the spatial coordinates. The KMS transformation of dynamical fields in fluid spacetime is given by
\begin{align}
    \widetilde{X}^\mu_a(-\sigma)=-X^\mu_a(\sigma) - \ii \beta_0\p_t X^\mu(\sigma)+\ii \beta_0 \delta^\mu_0,\;\;
    \widetilde{\varphi}_a (-\sigma) = -\varphi_a(\sigma)-\ii \beta_0 \p_t \varphi(\sigma),\;\; \widetilde{\varphi}^b_a (-\sigma) = \varphi^b_a(\sigma)+\ii \beta_0\p_t \varphi^b(\sigma),
\end{align}
and that of external gauge fields is given by
\begin{align}
    \widetilde{e}^b_{a,\mu}(-\sigma) = e^b_{a,\mu}(\sigma)+\ii \beta_0 \p_t e^b_\mu(\sigma), \;\; \widetilde{A}_{a,\mu}(-\sigma) = A_{a,\mu}(\sigma)+\ii \beta_0 \p_t A_\mu(\sigma),\;\; \widetilde{A}^b_{a,\mu}(-\sigma) = -A^b_{a,\mu}(\sigma)-\ii \beta_0 \p_t A^b_\mu(\sigma).
\end{align}
In the classical limit and physical spacetime, we thus have
\begin{subequations}\label{eq:mainKMS}
\begin{align}
    \widetilde{e}^b_{\mu}(-x) & = e^b_{\mu}(x),\\
    \widetilde{E}^b_{a,\mu}(-x) & = E^b_{a,\mu}(x)+\ii \CL_\beta e^b_\mu (x), \\
    \widetilde{B}_\mu (-x) & =  B_\mu(x),\\
    \widetilde{C}_{a,\mu}(-x) &= C_{a,\mu}(x)+\ii \CL_\beta B_\mu (x),\\
    \widetilde{K}^b_{\mu}(-x) &= -K^b_{\mu}(x),\\
    \widetilde{K}^b_{a,\mu}(-x) &= -K^b_{a,\mu}(x)-\ii \CL_{\beta} K^b_{\mu}(x),
\end{align}
\end{subequations}
where $\beta^\mu\equiv \beta_0 u^\mu$, and
\begin{subequations}
\begin{align}
    \CL_\beta e^b_\mu &= \beta_0\p_\mu u^b,\\ \CL_\beta B_\mu &= \beta_0\p_\mu  \mu+\beta^\nu(F_{\nu\mu}+2 e^b_{[\mu}\p_{\nu ]}\varphi_b),\\
    \CL_{\beta} K^b_{\mu}&= \p_\mu\left( \beta^\nu(\p_\nu \varphi^b+A^b_\nu)\right)+\beta^\nu F^b_{\nu\mu}.
\end{align}
\end{subequations}

\subsection{Ideal hydrodynamics}\label{sec:ideal}

To describe ideal hydrodynamics, it is convenient to first introduce the single-time equilibrium action \cite{jensenwithoutentropy}:
\begin{align}
    S_{0} = \int \mathrm{d}^{d+1}x  e P(e^b_t,B_t,K^b_t, K^b_I).
\end{align} 
To preserve rotational symmetry, the tensorial variables must be contracted by invariant tensors.
Then, the factorizability condition leads to
\begin{subequations}\label{eq:idealL}
\begin{align}
    & I_{\mathrm{EFT,eq}} = \int \mathrm{d}^{d+1}x \CL_{\mathrm{eq}} = S_0[\Lambda_1] - S_0[\Lambda_2], \\
    & \CL_{\mathrm{eq}} = p e^\mu_b E^b_{a,\mu}+n u^\mu C_{a,\mu}+\hat{\pi}_b u^\mu E^b_{a,\mu}+\psi^{\mu}_b \left(K^b_{a,\mu} - E^c_{a,\mu}e^\nu_c K^b_\nu\right),
\end{align}
\end{subequations}
where for the last term we used $(K^b_A e^A_\beta)_a = K^b_{a,\mu}e^\mu_\beta-K^b_\nu e^\nu_c e^\mu_\beta E^c_{a,\mu}$.
The coefficients in $\CL_{\mathrm{eq}}$ define the equation of state,
\begin{align}\label{eq:eosmain}
    p\equiv P, \quad n = \beta_0\frac{\p P}{\p B_t},\quad \hat{\pi}_b = \beta_0 \frac{\p P}{\p e^b_t},\quad \psi^{\mu}_b = \frac{\p P}{\p K^b_\mu},
\end{align}
where the partial derivatives of thermodynamic pressure $P$ are taken with other arguments being fixed. It can be verified that $\CL_{\mathrm{eq}}$ satisfies the KMS condition.
Now, we can read off the equilibrium stress tensor and currents from varying the action with respect to $e^b_{a,\mu}$, $A_{a,\mu}$ and $A^b_{a,\mu}$:
\begin{subequations}\label{eq:idealTJ}
\begin{align}
    T^\mu_{(0)b} &=p e^\mu_b +\pi_b u^\mu - \psi^\mu_c K^c_\nu e^\nu_b , \\
    J^\mu_{(0)} &= n u^\mu,\\
    J^{\mu}_{(0)b} & = \psi^{\mu}_b,
\end{align}
\end{subequations}
where we defined the momentum density as
\begin{align}\label{eq:momdensity}
    \pi_b \equiv \hat{\pi}_b + n\varphi_b \equiv \hat{\rho}u_b+n\varphi_b,
\end{align}
with $\hat{\rho}\geq 0$ an $\CO(1)$ coefficients.
We can further express the dipole currents by expanding the pressure up to quadratic terms in field amplitude
\begin{align}
    P \sim \frac{1}{2} b K_{0b}K_{0b} - \frac{1}{2} a^{ibjc} K_{ib} K_{jc}+\ldots,
\end{align}
where the invariant tensor is given by
\begin{align}\label{eq:atensor}
    a^{ijkl} &= a_1 \delta^{ij}\delta^{kl}+a_2 \delta^{i<k} \delta^{l>j} +2a_3\delta^{i[k} \delta^{l]j},
\end{align}
with $A^{<ij>}=A^{ij}+A^{ji}-\frac{2}{d}\delta^{ij}A^{kk}$, $A^{[ij]} = \frac{1}{2}(A^{ij} - A^{ji})$.
Thermodynamic stability requires that $b,a_{1,2,3}\geq 0$. 
Thus, we have 
\begin{subequations}
\begin{align}
    J^{0}_{(0)b}& = \psi^0_b =    b K_{0b}, \\
    J^{i}_{(0)b}& = \psi^i_b =   -  a^{ibjc} K_{jc}.
\end{align}
\end{subequations}
Let us turn off the gauge field temporarily and plugin \eqnref{eq:idealTJ} into \eqnref{eq:WImain}.  Then, we find exactly the momentum and charge conservations, but with an additional dipole constraint:
\begin{align}\label{eq:uvarphi}
    n u_b =  - a^{ibkc}\p_i\p_k \varphi_c.
\end{align}
The above vanishes on equilibrium, which can be viewed as a consequence of the dipole and momentum algebra.
Here, we have neglected the terms with time derivatives; in fact, as we will see in \secref{sec:dissipation}, there is a relaxation time that relaxes non-hydrodynamic modes, which makes (\ref{eq:uvarphi}) exact (up to a fluid frame change).
As a result, \eqnref{eq:momdensity} reduces to
\begin{align}\label{eq:pi}
    \pi_b = n\varphi_b + \cdots,
\end{align}
up to higher order corrections. The low energy excitations for the dipole-conserving fluid are not single particle (fracton) excitations, but rather propagating dipole Goldstone/density waves. In particular, the momentum susceptibility $\rho$, defined as $\pi_b \equiv \rho u_b$ (different from $\hat{\rho}$), will be non-local:
\begin{align}\label{eq:momsusc}
    \rho\sim \frac{n^2}{k^2},
\end{align}
and diverges at large distance $k\to 0$; see \appref{sec:MM} for another derivation of non-local $\rho$ using the memory matrix formalism.

At leading order in amplitude expansion, the conservation equations for $\delta n\equiv n-n_0$ and $\pi_b$ are
\be \label{eq:eqeom}
\p_0 \pi_b +\frac{n_0}{\chi}\p_i\delta n \delta^i_b=0,\qquad \p_0 \delta n-\frac{a}{n_0} \p_i^2\p_j \pi_b\delta^{jb}=0\,,\ee
where we only retained terms at leading order in derivatives and $a = a_1+2a_2(d-1)/d $. Upon Fourier transformation, the normal modes are given by
\begin{align}\label{eq:magnonideal}
    \omega = \pm \sqrt{\frac{a}{\chi}} k^2.
\end{align}
We therefore find that the dipole ``sound'' modes are magnon-like.

\subsection{Dissipative hydrodynamics and higher-order terms}\label{sec:dissipation}

We are now ready to write down the most general (leading order) dissipative part of the effective field theory. We expand the Lagrangian to containing at most two factors of $a$-fields
\begin{align}\label{eq:Lwoenergy}
    \CL = \CL_{\mathrm{eq}}+\CL^{(1)}+\overline{\CL}^{(1)}+\CL^{(2)},
\end{align}
where the superscript $(n)$ represents the number of $a$-fields, and we will always keep the leading derivative orders. As mentioned around \eqref{eq:mainKMS}, we will restrict to systems whose microscopic dynamics is $\CP\CT$-even dynamics. The dynamical KMS symmetry then implies that $\CL^{(2)}$ is $\CP\CT$-even and relates it to $\CL^{(1)}$; moreover it allows the presence of an additional $\overline{\CL}^{(1)}$ that is KMS-invariant by itself (and is thus also $\CP\CT$-even).
It is helpful to first introduce a combined field
\begin{align}
    \Delta U_{a,ib} &\equiv   \p_i C_{a,j}e^j_b -K_{a,ib}\nonumber\\ 
    &= \p_i\p_j \varphi_a e^j_b +\p_i(A_{a,j}+\CL_{X_a}A_j)e^j_b +\p_i(E^c_{a,j}\varphi_c)e^j_b -( A_{a,i b}+\CL_{X_a}A_{i b}),
\end{align}
whose KMS transformation is given by
\begin{align}
    \widetilde{\Delta U}_{a,ib} (-x) =  -\Delta U_{a,ib} (x) - \ii \CL_{\beta} \Delta U_{ib}(x),
\end{align}
where
\begin{align}
    \CL_{\beta} U_{ib} &\equiv  \beta_0e^j_b \p_i\p_j \mu +\p_i(\beta^\nu F_{\nu j})e^j_b +\p_i(\beta^c\p_j \varphi_c)e^j_b  -\p_i(\beta^\nu A^b_\nu) - \beta^\nu F^b_{\nu i} \nonumber \\
    &\approx \beta_0e^j_b \p_i\p_j \mu+\beta_0\p_i F_{0j}e^j_b - \beta_0 \p_0 A^b_i,
\end{align}
with the second line being the approximation of linear response.
With the benefit of hindsight, we have constructed this field to be independent of $\varphi^b_a$.
Using this field, we can write the most general $\CP\CT$-even $\CL^{(2)}$ as
\begin{align}\label{eq:L2}
    -\ii \beta_0 \CL^{(2)} =\sigma^{ij} C_{a,i}C_{a,j}+ s^{ibjc}E_{a,i b} E_{a,j c}+t^{ibjc}\Delta U_{a,ib}\Delta U_{a,jc}+2r^{ijkb}_1\p_i C_{a,j}\Delta U_{a,kb}+r^{ijkl}_2\p_i C_{a,j}\p_k C_{a,l},
\end{align}
where $\sigma^{ij} = \sigma\delta_{ij}, \sigma\geq 0$.
In the above action, we only considered  leading derivative contributions from $a$-type fields, and we additionally included first derivative terms in $C_{a,i}$ as they are of the same order as $\Delta U_{a,ib}$. As usual, the terms proportional to $E^b_{a,0}$, $C_{a,0}$ can be eliminated by field redefinition as shown below, and, at the same time, $K^b_{a,0}\sim \p_t\varphi^b$ is neglected as it is subleading to the last term which contains first spatial derivatives of $\varphi^b$, and we are keeping into account the scaling $\omega\sim k^2$ found in \eqnref{eq:magnonideal}. Under KMS transformations, we require $[\CL^{(2)}+\CL^{\prime(1)}- (\tilde{\CL}^{(2)}+\tilde{\CL}^{\prime(1)})]_{\CO(a)}=0$, where $\CO(a)$ indicates the (first) order of $a$-fields. Since $\CL^{(2)}|_{\CO(a)}=0$, the constraint reduces to $\CL^{\prime(1)}-\tilde{\CL}^{\prime(1)}|_{\CO(a)}=\tilde{\CL}^{(2)}|_{\CO(a)}$. Next, we redefine $\CL^{(1)} \equiv \frac{1}{2}( \CL^{\prime(1)}-\tilde{\CL}^{\prime(1)})_{\CO(a)} = \frac{1}{2}\tilde{\CL}^{(2)}|_{\CO(a)}$, thus, by construction, $\CL^{(1)}$ is $\CP\CT$-odd.
This leads to
\begin{align}\label{eq:L1}
    \beta_0\CL^{(1)} =& - \sigma^{ij} \CL_\beta B_i C_{a,j} -s^{ibjc} \CL_{\beta}e_{ib} E_{a,jc} - t^{ibjc}\CL_{\beta}\Delta U_{ib} \Delta U_{a,jc} - r^{ijkb}_1 \p_i\CL_\beta B_j \Delta U_{a,kb} - r^{kbij}_1 \CL_{\beta} \Delta U_{kb}\p_i C_{a,j}\\ \nonumber
    & - r^{ijkl}_2 \p_i\CL_\beta B_j \p_k C_{a,l} .
\end{align}
From the above discussion, the effective Lagrangian allows a $\CP\CT$-even $\overline{\CL}^{(1)}$ that itself remains invariant under KMS transformation. This is given by
\begin{align}
    \beta_0\overline{\CL}^{(1)} = d^{ibjc}\left(\CL_{\beta}e_{ib}  \Delta U_{a,jc} - E_{a,ib} \CL_{\beta}\Delta U_{jc} \right)+f^{ijkb} \left(\CL_{\beta}e_{kb}  \p_i C_{a,j} - E_{a,kb}\p_i \CL_{\beta}B_{j} \right),
\end{align}
which describes non-dissipative dynamics.
So far, the effective field theory is general and complete, but we will see below that simplifications can be made by ignoring certain higher order corrections. 

From \eqnref{eq:uvarphi}, we see that the dipole Ward identity is not a conservation law but a force balance equation. We now show that the associated field $\varphi^b_{a}$ can be eliminated and still preserve locality of the effective action. Indeed, by integrating out $X^i$ in \eqnref{eq:idealL}, we obtain
\begin{align}\label{eq:intXi}
    0 = \p_0\left[ C_{a,i}+\frac{\hat{\rho}}{n_0}E_{a,0b}e^b_i+\cdots \right]=\p_0\left[ \p_i \varphi_a+A_{a,i}+\CL_{X_a}A_i+\varphi_{a,b}e^b_i+E^b_{a,i}\varphi_b+\frac{\hat{\rho}}{n_0}E_{a,0b}e^b_i+\cdots \right] ,
\end{align}
where the dots include higher derivative orders. As all the fields are set to be zero at spacetime infinity, the expression in the bracket is also zero, and since $\varphi^b_{a}$ appears without derivatives we can eliminate it from the effective action without generating non-local terms. Importantly, the combined field $\Delta U_{a,ij}$ does not contain $\varphi^b_a$, so we simply need to replace $C_{a,i}$:
\be
C_{a,i}=-\frac{\hat{\rho}}{n_0}E_{a,0b}e^b_i+\cdots\,.
\ee
After such replacement, the possible additional effective Lagrangian can be added is
\begin{align}\label{eq:additionL}
    \beta_0\CL^{(1)}_{\mathrm{addition}}\sim -A \CL_\beta B_0 C_{a,0} - B \CL_{\beta} e_{0,b}E_{a,0b}.
\end{align}
Now, suppose that we are able to shift the $r$-fields through\footnote{Since $X^i$ has been integrated out, $u^\mu$ is not a low-energy degree of freedom to which the field redefinition can be applied.}
\begin{align}
    \mu \to \mu+\delta \mu,\quad \varphi^b\to \varphi^b+\delta \varphi^b,
\end{align}
then the correction to $\CL^{(1)}$ from $\CL_{\mathrm{eq}}$ is given by
\begin{align}
    \delta_r \CL_{\mathrm{eq}}\sim \delta n_0 C_{a,0}+n_0 \delta\varphi_b E_{a,0b},
\end{align}
where $\delta n_0 = \delta\mu \p_\mu n_0+\delta \varphi^b \p_{\varphi^b}n_0$, and we have neglected the contribution to the bulk viscosity. We find that if we choose the field redefinition as 
\begin{align}
    \delta n_0 = \beta_0^{-1}A\CL_{\beta}B_0,\quad \delta\varphi^b = (n_0\beta_0)^{-1} B \CL_{\beta}e_{0,b},
\end{align}
then the additional Lagrangian \eqnref{eq:additionL} can be eliminated. This indicates that terms proportional to $C_{a,i}$, $\p_iC_{a,j}$ can be safely ignored as a change of frame, and the effective Lagrangian becomes
\begin{subequations}\label{eq:newL2}
\begin{align}
    -\ii \beta_0 \CL^{(2)} &= s^{ibjc}E_{a,i b} E_{a,j c}+t^{ibjc}\Delta U_{a,ib}\Delta U_{a,jc},\\
    \beta_0\CL^{(1)} &=  -s^{ibjc} \CL_{\beta}e_{ib} E_{a,jc} - t^{ibjc}\CL_{\beta}\Delta U_{ib} \Delta U_{a,jc},\\
    \beta_0\overline{\CL}^{(1)} &= d^{ibjc}\left(\CL_{\beta}e_{ib}  \Delta U_{a,jc} - E_{a,ib} \CL_{\beta}\Delta U_{jc} \right).
\end{align}
\end{subequations}

In parallel, if we do not integrate out $X^i$, we find that the coefficient $\sigma$ associated with $C_{a,i}$ in \eqnref{eq:L2} gives the relaxation of a non-hydrodynamic mode. By varying \eqnref{eq:L1} with respect to $\varphi^b_a$, we obtain the leading-order dipole Ward identity
\begin{align}\label{eq:relaxeom}
    b \p_0^2 \varphi_b - a^{ibjc} \p_i\p_j\varphi_c  = n u_b - \sigma\left(e^i_b\p_i\mu+\p_0\varphi_b\right).
\end{align}
Clearly, $\p_0\varphi_b$ acquires a relaxation time $\tau$: 
\begin{align}\label{eq:relaxation}
    \tau \equiv \frac{b}{\sigma}.
\end{align}
Therefore, on a finite time scale $\tau$, $\partial_0\varphi_b$ relaxes to $u_b$ (schematically) -- thus they are not independent degrees of freedom in our hydrodynamic limit ($t\rightarrow \infty$):  $\varphi_b$ is the hydrodynamic mode corresponding to momentum density.
We thus understand that $\sigma$ is not the usual transport coefficient but determines the relaxation rate for the dipole Ward identity to become a force balance equation. This allows us to ignore it in the hydrodynamic limit.

Hence, we see that $X^i$ is not necessarily a physical degree of freedom (at finite density).  What is non-trivial is that the momentum density, which ordinarily would be $\partial_0X^i$, is approximately proportional to the dipole Goldstone.  It seems non-trivial to uncover the ultimate structure we have found without introducing these extra degrees of freedom, but it may be possible to achieve this in future work.  In particular, we find it most instructive to couple this theory to geometry (see \appref{app:curved}) in the presence of such additional degrees of freedom.

The derivative expansion of the stress tensor and currents are obtained from variation of $\CL^{(1)}+\overline{\CL}^{(1)}$, which leads to (neglecting nonlinear terms, which will not be relevant for the remainder of this section)
\begin{subequations}\label{eq:first-order}
\begin{align}
    T^{ib}_{(1)} & = -s^{ibjc}\p_j u_c - d^{ibkl} \left(\p_k\p_l\mu+\p_kF_{0l}   - A_{kc}\delta^c_l\right),\\
    J^{ib}_{(1)} & =t^{ibkl}\left(\p_k\p_l\mu+\p_kF_{0l} - A_{kc}\delta^c_l\right)- d^{jcib} \p_j u_c,
\end{align}
\end{subequations}
where $u_b$ is fixed by \eqnref{eq:uvarphi}.  In a rotationally invariant theory, we have
\begin{subequations}\label{eq:tensors}
\begin{align}
    s^{ijkl} & = \zeta \delta^{ij}\delta^{kl}+\eta \delta^{i<k}\delta^{l>j} ,\\
    t^{ijkl} &= t_{1}\delta^{ij}\delta^{kl}+t_{2} \delta^{i<k}\delta^{l>j}\\
    d^{ijkl} & = d_1 \delta^{ij}\delta^{kl}+d_2 \delta^{i<k}\delta^{l>j}.
\end{align}
\end{subequations}
From the unitarity of the effective action, $\mathrm{Im}\CL^{(2)}\geq 0$, we find that the dissipative coefficients satisfy the following positivity constraint,
\begin{align}
    \zeta,\eta,t_1,t_2\geq 0,
\end{align}
while $d_{1,2}$ are unconstrained and non-dissipative.

Let us now analyze the normal modes around a homogeneous background charge density $n_0>0$. We also turned off the background fields for simplicity. Treating the deviation $\delta n=n-n_0$ and $\varphi_b$ as small, we obtain the derivative expansion of the pressure as
\begin{align}
    p&\approx p_0 + \frac{\p P}{\p B_t}|_{K^b_\mu, e^i_t}\delta B_t + \frac{1}{2}\frac{\p P}{\p K^b_i}|_{B_t, e^i_t} \p_i \varphi^b +\ldots,\\\nonumber
    &\approx p_0+ \left[ \chi^{-1}n_0\delta n+ \frac{1}{2} \frac{\p^2 p_0}{\p n_0^2}(\delta n)^2 \right]   -\frac{1}{2} \left(a^{ibjc} +\frac{\p a^{ibjc}}{\p n_0} \delta n \right) \p_i\varphi_b \p_j\varphi_c +\ldots,
\end{align}
where $\chi = \frac{\p n}{\p \mu}$ is the normal charge susceptibility.
Hence, the stress tensor and currents are given by
\begin{subequations}\label{eq:nonlTJ}
\begin{align}
    T^{ib}&\approx \left(p_0+\chi^{-1}n_0 \delta n+\frac{1}{2}\frac{\p^2 p_0}{\p n_0^2}(\delta n)^2-\frac{1}{2}a^{kdjc}\p_k\varphi_d\p_j\varphi_c\right)\delta^{ib} \\\nonumber
    &\;\;\;\;-a^{jikc}\varphi^b \p_j\p_k\varphi_c+a^{icjd}\p_j\varphi_d\p_k\varphi_c \delta^{kb}+T^{ib}_{(1)}+\tau^{ib},\\
    J^{ib}&\approx -\left(a^{ibjc}+\frac{\p a^{ibjc}}{\p n_0}\delta n\right)\p_j\varphi_c +J^{ib}_{(1)}+\xi^{ib},
\end{align}
\end{subequations}
where $\tau^{ib},\xi^{ib}$ are the noise whose variance satisfy the fluctuation-dissipation theorem:
\begin{subequations}\label{eq:noise}
\begin{align}
    \langle\tau^{ib}(x)\tau^{jc}(0) \rangle &= 2T_0 s^{ibjc} \delta^{(d+1)}(x),\\
    \langle\xi^{ib}(x)\xi^{jc}(0) \rangle &= 2T_0 t^{ibjc} \delta^{(d+1)}(x).
\end{align}
\end{subequations}
The expressions for $T^{ib}_{(1)}$ and $J^{ib}_{(1)}$ are given by \eqnref{eq:first-order} with all the coefficients taking their equilibrium values; we can neglect the non-dissipative terms proportional to $d_{1,2}$ since they are sub-leading corrections to the ideal hydrodynamics.
So far, we have included the leading order derivatives and only kept non-linear terms in the ideal hydrodynamics because the non-linearity in dissipative coefficients are irrelevant \cite{Glorioso:2021bif}. Plugging \eqnref{eq:nonlTJ} in \eqnref{eq:WImain} and using \eqnref{eq:uvarphi} and \eqnref{eq:pi}, we obtain equation of motions:
\begin{subequations}\label{eq:eommom}
\begin{align}
    n_0\p_0 \varphi^b+\chi^{-1}n_0 \p_i \delta n \delta^{ib}+\lambda n_0 \delta n \p_i \delta n \delta^{ib}+2a^{icjd}\p_i\p_j \varphi_d \p_{[k} \varphi_{c]} \delta^{kb} 
    +n_0^{-1} s^{ibjc}a^{kcld}\p_i\p_j\p_k\p_l \varphi_d + \p_i\tau^{ib}&=0, \\
    \p_0 \delta n-a^{ijkb}\p_i\p_j\p_k \varphi_b- \bar{\lambda}^{ijkb}  \p_i\p_j(\delta n \p_k \varphi_b)+\chi^{-1} t^{ijkl}\p_i\p_j\p_k\p_l \delta n+\p_i\p_j \xi^{ib} \delta^j_b &=0,
\end{align}
\end{subequations}
where we denoted $\lambda = n_0^{-1}\p_{n_0}^2p_0$, $\bar{\lambda}^{ijkb} = \p_{n_0}a^{ijkb}$ as the non-linear coefficients. In the above equation, we have neglected the higher order time derivatives and assumed that all the coefficients upon expansion do not depend on $\delta n$ and $\varphi_i$. 
The normal modes are defined as non-vanishing solutions to the equation of motion. By neglecting the non-linear terms, we obtain
\begin{subequations}\label{eq:normalmom}
\begin{align}
    \omega &= \pm \sqrt{\frac{a}{\chi}} k^2 - \ii \left(\frac{t}{\chi}+\frac{1}{n_0^2}(\Gamma_1+\Gamma_2)\right)k^4,\label{eq:magnon}\\
    \omega&=-\ii \frac{\Gamma_1}{n_0^2} k^4,\label{eq:subdiff}
\end{align}
\end{subequations}
where
\begin{align}\label{eq:defnm}
    a &= a_1+2\frac{d-1}{d}a_2,\nonumber\\
    t &= t_1+2\frac{d-1}{d}t_2,\nonumber\\
    \Gamma_1 &= (a_2+a_3)\eta ,\nonumber \\
    \Gamma_2 &= \zeta a_1+2\frac{d-1}{d}(\zeta a_2+\eta a_1)+\left(3-\frac{8}{d}+\frac{4}{d^2}\right)\eta a_2 - \eta a_3.
\end{align}
Therefore, we find two longitudinal propagating mode with magnon-like dispersion relation $\sim\pm k^2$ and attenuation $\sim -\ii k^4$ in \eqnref{eq:magnon}, and transverse subdiffusive modes $\sim -\ii k^4$ with multiplicity $d-1$ in \eqnref{eq:subdiff}.

\subsection{Relevant perturbations in low dimensions}

Following \cite{Glorioso:2021bif}, we give a zeroth-order scaling analysis on the nonlinear effect. The dissipative scaling $\omega\sim k^4$ causes the noise to scale as $\tau^{ib},\xi^{ib}\sim k^{(d+4)/2}$ based on \eqnref{eq:noise}. To match the scaling to the dynamical terms with time derivatives, we find $\varphi^b\sim k^{d/2-1}$ and $\delta n\sim k^{d/2}$. As per the usual renormalization group analysis, the nonlinear coefficients would scale as $\lambda, \bar{\lambda}^{ijkb}\sim k^{(4-d)/2}$, making them relevant when $d<4$. As a consequence, we expect the true IR fixed point to have anomalous dissipative scaling: $\omega\sim \pm k^2-\ii k^z$, with $z<4$. We crudely estimate $z$ by assuming that the thermodynamic field does not renormalize due to its Gaussian fluctuations at long wavelength. Then, requiring $\lambda, \bar{\lambda}^{ijkb}\neq 0$ to not depend on $k$ at fixed point, we obtain $z=d/2+2$. We emphasize that the critical exponent $z$ has been testified numerically in \cite{Glorioso:2021bif} with excellent agreement in $d=1,2$, suggesting a breakdown of hydrodynamics.
Meanwhile, naive scaling analysis also implies that the transverse subdiffusive modes would acquire an anomalous scaling $\omega\sim -\ii k^{z'}$ with $z'=z$. We hope to report on a 1-loop analysis of the corrections to hydrodynamics in the near future in order to further investigate these claims.  

Interestingly, there was previous work on a stochastic molecular-beam-epitaxy (MBE) process \cite{Sarma1991} which shares some similarity with our model. In \cite{Sarma1991}, the authors study \begin{equation}
\partial_t h + \nabla^2\left(\nabla h \right)^2 + \nabla^4h + \text{noise}=0.
\end{equation}
This equation, like our theory, has $z=4$ subdiffusion at the linear level.  However, the authors of \cite{Sarma1991} did not demand invariance of the renormalized theory under dipole shift $h\rightarrow h+c_0+c_1x$, and as such they argued that while (like us) the critical dimension $d=4$, they instead found a distinct fixed point with $z=(8+d)/3$, which we understand as coming from fixing the scaling dimension of noise, rather than of $h$.  It may be possible to interpret the results of \cite{Sarma1991}, in light of our work, as highlighting the possibly ``accidental" appearance of the same dynamical fixed point (KPZ) in both a hydrodynamic setting (Navier-Stokes equations in 1d), and as a model of surface growth.   As our symmetries and anticipated scaling exponents differ from \cite{Sarma1991}, the analogy between surface growth and hydrodynamics with momentum conservation may break down in multipole-conserving theories.

\section{Spontaneous symmetry breaking}\label{sec:ssb}
In this section, we discuss a subtle yet very interesting feature of the dipole and momentum conserving fluid: the identification of $\varphi_b$ as a Goldstone boson of the dipole field, and the corresponding intuition that dipole symmetry is (by many reasonable definitions) \emph{spontaneously broken}.

Let us first briefly justify the identification of $\varphi_b$ as a Goldstone boson.  Under dipole shift symmetry, $\varphi_b \rightarrow \varphi_b + c_b$, just as a conventional Goldstone.  Moreover, if we wanted to consider (even in the absence of momentum) the spontaneous symmetry breaking of the dipole charge \cite{stahl}, this could be achieved by adding $\nabla \varphi_b \nabla \varphi_{a,b}$ terms to the Lagrangian (this will be discussed further elsewhere).  Recent papers on dipole symmetry breaking in the absence of momentum conservation include \cite{Peng_2020prr,Lake:2022ico}, while \cite{Kapustin:2022fzp} discusses the role of dipole symmetry breaking in a translation invariant state.   Lastly, as we will see, the dynamics of the collective and long-lived $\varphi_b$ mode saturate Goldstone's Theorem.  For these three reasons, we will call $\varphi_b$ the Goldstone boson associated with dipole symmetry.

There are three notions through which prior authors have justified the presence of spontaneous symmetry breaking that we are aware of. Briefly, and we will elaborate more on each in subsequent sections: (\emph{1}) The physical state of interest is not invariant under the action of the global symmetry group; (\emph{2}) the existence of a Goldstone boson which saturates Goldstone's Theorem; (\emph{3}) the presence of long-range order in expectation values of operators that shift under the symmetry (Mermin-Wagner Theorem).  In a nutshell, the theory of interest here is compatible with all three definitions in spatial dimensions $d>2$, but only with the first two when $d \le 2$.  In low dimensions, our theory thus seems to represent an unusual paradigm not encountered before, where many (but not all) of the usual features of SSB exist. We believe that it is appropriate to consider dipole symmetry as spontaneously broken, but leave it to future authors to more firmly settle the possibly semantic question.   We conclude this section with a discussion of a quantized version of Model A from \cite{Glorioso:2021bif}, which will give a concrete example of the more abstract ideas discussed.

\subsection{Mermin-Wagner Theorem}\label{sec:mwt}
First, we focus on a physically transparent test for spontaneous symmetry breaking:  correlation functions of the order parameter $\pi^i(t,x)$.   A microscopic argument along these lines was presented in \cite{Kapustin:2022fzp}, and previously in \cite{stahl} in the absence of momentum conservation. 
While this operator is charged under dipole transformations, it transforms nonlinearly: $\pi^i\to \pi^i + c^i$, where $c^i$ is an arbitrary vector parameter.\footnote{In this subsection we will be agnostic of the global structure of the dipole group and of possible subtleties related to boundary conditions. We will describe these in more detail for a specific model in the next subsection.} To conclude that dipole symmetry is spontaneously broken, one often asks whether $\pi^i$ is a well-defined order parameter: any finite value (including zero) will be sufficient to conclude spontaneous breaking. From the discussion around (\ref{eq:momsusc}), the equal-time two-point function of $\pi^i$ diverges at low momenta: \begin{equation}
    \langle \pi^i(k)\pi^j(-k)\rangle\sim \frac 1{k^2}.
\end{equation} Let us consider the average momentum density on a region of linear size $L$. The fluctuations of the average momentum density on this region scale as
\begin{equation}
    \left\langle (\pi^i - \langle \pi^i\rangle)^2\right\rangle_L\sim \left\lbrace \begin{array}{ll} L &\ d=1 \\ \log L &\ d=2 \\ \text{constant} &\ d>2 \end{array}\right..
\end{equation} This means that the momentum density is well-defined as a thermodynamic variable only for $d>2$, in which case (\ref{ssb1}) implies spontaneous symmetry breaking of dipole transformations. For $d\leq 2$, fluctuations are too large to make sense of the expectation value of $\pi^i$, and therefore we cannot conclude that there is spontaneous symmetry breaking from this perspective.  

\subsection{Goldstone's Theorem}
Let us now discuss how the classic Goldstone's Theorem can be generalized to our theory.  Consistency with the algebra (\ref{eq:DPQ}) demands the following commutation relation between dipole and momentum density:
\be [D_i,\pi^j]=in\delta^{j}_i\,.\ee
On a thermal state $\rho_0$ at finite background charge density $n_0$, we have
\be \label{ssb1}\tr(\rho_0[D_i,\pi_j])=i n_0 \delta_{ij}\neq 0\,.\ee
For concreteness, we shall take $\rho_0$ to be microcanonical with respect to momentum and charge, and canonical with respect to energy (if the latter is conserved); see an explicit example in \secref{sec:modelA}. As we discussed in the previous section, $\pi^i$ has large fluctuations in low dimensions; nevertheless the expression on the left-hand side of (\ref{ssb1}) can still be well-defined as we demonstrate in the explicit example of  \secref{sec:modelA}. We now show how this relation \eqnref{ssb1}, entirely dictated by symmetry, will imply  the existence of a Goldstone mode \cite{nicolisSSB}.
Let us start by writing the total dipole charge in terms of the charge density operator $D_i=\int d^d x\, x^i n$,\footnote{The total dipole moment $D_i$ could in principle receive further contributions from ``bond dipoles'', i.e. degrees of freedom with an internal dipole charge. This is entirely analogous to the contribution of orbital angular momentum and spin to the total angular momentum. We shall not investigate this situation here; see hydrodynamics of spin in \cite{hattori2019fate,Gallegos:2021bzp,Gallegos:2022jow}. Note, however, that these bound dipole degrees of freedom are not conserved and have a finite lifetime. Therefore it is unlikely they would contribute to the singular spectral weight we calculate below.} which allows us to express the above as
\be \int d^d x\, x^i G^R_{n\pi_j}(t,x)=- n_0\theta(t) \delta^{ij},\qquad G^R_{n\pi_i}(t,x)=-i\theta(t)\tr(\rho_0[n(t,x),\pi_i(0,0)])\,,\ee
where $G^R_{n\pi_i}$ is the retarded two-point function of charge and momentum densities. Doing a Fourier transform leads to
\be\label{satz} \lim_{\vec p\to 0}\frac{\p}{\p p^i} \text{Im}G^R_{n\pi_i}(\omega,p)= -n_0\pi\delta(\omega)\delta_{ij}\,.\ee
We see that \eqnref{satz} implies a zero-frequency contribution to the spectral density $J_{n\pi^i}$ as a direct consequence of \eqnref{ssb1}. 

Using the approach of Kadanoff-Martin and \eqnref{eq:eqeom}, we now show that this spectral weight is entirely captured by a single hydrodynamic mode: the ``magnon-like" sound mode.  We can obtain the retarded two-point function $G^R_{n\pi_i}$ (without missing any important counterterm). Doing a Laplace transform of \eqnref{eq:eqeom},
\be \begin{pmatrix} -iz \delta_{ij} & \frac{n_0}\chi i k_i\\ ik_j k^2\frac{a}{n_0}& -i z\end{pmatrix}\begin{pmatrix}
\pi_j(z,p)\\\delta n(z,p)\end{pmatrix}=\begin{pmatrix}\pi^{(0)}_i(p)\\ \delta n^{(0)}(p)\end{pmatrix}\,,\ee
where the right-hand side represents densities configurations at the initial time. The initial density configuration $\delta n^{(0)}$ is related to a perturbation in the chemical potential at initial time through $\delta n^{(0)}=\chi\delta\mu$, so we find
\be \frac{\p\pi_i(z,p)}{\p \delta \mu(p)}=\frac {-n_0 i k_i}{-z^2+\frac{a}{\chi} k^4}\,.\ee
From linear response we know that, for a density $\delta n_a$ conjugated to a chemical potential $\delta \mu_a$, $iz\delta n_a(z,k)=(G^R_{ab}(z,k)-G^R_{ab}(0,k))\delta \mu_b(k)$ and $G_{ab}^R(0,k\to 0)=-\chi_{ab}$. We then find
\be G^R_{n\pi_i}(\omega,p)=\frac{-n_0 \omega k_i}{-\omega^2+\frac{a}{\chi} k^4}\,.\ee
At leading order in $k_i$, we have
\be \text{Im}\, G^R_{n\pi}(\omega,p)\to -\pi n_0 k_i\delta(\omega)\ee
which can be obtained by taking into account that dissipative corrections in the retarded two-point fuction contribute through $\omega\to \omega+i\vep$, where $\vep\to 0$ as $k\to 0$. We then see that we recovered eq. (\ref{satz}). This contribution comes from the pressure term, and is thus present in any fluid which conserves momentum and density. 

Hence we call $\varphi_b$ a Goldstone mode: this IR mode by itself saturates the requisite Goldstone's Theorem for dipole symmetry.   The fact that in the EFT, $\varphi_b$ did not have a diagonal shift symmetry like $\varphi$ and $X_i$, further suggests that we should interpret $\varphi_b$ as a Goldstone mode.  In previous hydrodynamic effective field theories such as \cite{Crossley:2015evo,Glorioso:2017fpd}, when this diagonal shift symmetry is not present, the conclusion has been that the corresponding continuous symmetry is spontaneously broken.


At charge neutrality $n_0=0$, the Goldstone's theorem become trivial, and we will not have such dipole Goldstones.


\subsection{Existence of a symmetry-breaking  state}\label{sec:modelA}

Finally, we now show that a system with dipole and momentum symmetry can possess a symmetry-breaking ground state even in one and two space dimensions. 

The dipole and momentum algebra is isomorphic to the Heisenberg algebra of position and momentum (in a microcanonical ensemble where $Q$ is fixed).  In textbook quantum mechanics (and in models interest here, as discussed below), the physical Hilbert space contains no trivial representation of this algebra.  Therefore it is not possible to construct a state which is invariant under the symmetry group, and one can say that the symmetry must be broken, either explicitly or spontaneously.  

A physical example of another system that has a (sub)algebra isomorphic to dipole and momentum is a Galilean-invariant fluid.  This is slightly different because the commutator of $[H,D]\ne 0$ in the Galilean-invariant fluid.  Nevertheless, recent papers \cite{nicolisSSB, komargodski_boost} have argued that Galilean boosts are spontaneously broken -- in any dimension -- by the fluid rest frame.  We believe this is most succinctly justified by the argument in the previous paragraph. We discuss an interesting relation between our dipole fluids and the Galilean-invariant fluids in \secref{sec:gal}.

One challenge in showing whether the ground state is invariant or not, is that historically this was done by a careful consideration of finite volume regularization \cite{MW_1966}.  When compactifying space onto a toroidal lattice with finite length in each direction, the dipole conservation law is isomorphic to $\mathbb{Z}$, not $\mathbb{R}$.\footnote{We thank Nathan Seiberg for emphasizing the points in this paragraph.}  This regularization is not acceptable for us here since it qualitatively changes the nature of the symmetry group; we seek an alternative regularization that manifestly preserves the Heisenberg algebra.

Model A of \cite{Glorioso:2021bif} provides us with a concrete model which we can suitably regularize. The Hamiltonian for Model A is \begin{equation}
    H = \sum_{i=1}^{N-1} \left[\frac{(p_i-p_{i+1})^2}{2} + V(x_{i+1} - x_{i}) \right]
\end{equation}
with $[x_i,p_j] = \mathrm{i}\delta_{ij}$ conventional position and momentum operators. We can view this as a chain where each site $i$ possesses an infinite-dimensional Hilbert space. Alternatively, we can interpret this as a generalization of an atomistic Hamiltonian modelling a solid in one dimension, where $x_i$ and $p_i$ describe displacements and momenta.  Similar models (without dipole conservation) are known to capture hydrodynamics (and its breakdown) in one dimension \cite{Das_heat_2014}. The details of $V$ are not important for our discussion, though we would likely consider a function Taylor expanded about a minimum at argument $x=1$.   One can easily see that $H$ commutes with the dipole and momentum operators of the theory: \begin{equation}
    P = \sum_{i=1}^N p_i, \;\;\;\;\;\;D = \sum_{i=1}^N x_i.
\end{equation}
Note that this theory is in a microcanonical ensemble for charge: $Q=N$ is a fixed $c$-number. To see that the thermodynamics is well defined, we take $\rho_0= e^{-\beta H}$ restricted to (at the quantum level) a fixed wave function for the center of mass coordinate $D$: see (\ref{eq:comwof}) below; we also take lattice constant $a=1$. Upon changing variables to $s_i=p_i-p_{i+1}$, $r_i=x_i-x_{i+1}$, the Hamiltonian becomes $H=\sum_{i=1}^{N-1}\frac 12 s_i^2+V(r_i)$.
The classical partition function is then (up to an overall constant) 
$\int \prod_{i=1}^{N-1}ds_i dr_i e^{-\beta H}$. The partition function factorizes into manifestly convergent integrals, and is therefore finite.  In quantum mechanics, the partition function must also be finite.  We additionally notice that due to the commutator $[D,p_i]=\text i$, $\tr(\rho_0 [D,p_i])$ is finite, thus showing that the atomistic analogue of (\ref{ssb1}) is well-defined in the present model, despite the large fluctuations of momentum density discussed in Section \ref{sec:mwt}.


Let us now look in more detail at the quantum ground state of Model A.  This will be a wave function of the form \begin{equation}\label{eq:comwof}
    \psi_{\mathrm{gs}}(x_1,\ldots,x_N) = \psi_0(x_1-x_2,\ldots,x_{N-1}-x_N) \cdot \mathrm{e}^{\mathrm{i}k(x_1+\cdots + x_N)} \frac{\mathrm{e}^{-(x_1+\cdots + x_N)^2/2Na}}{(2\pi a)^{1/4}}.
\end{equation}
This wave function has an eigenvalue independent of $k$ and $a$. If we take $a=\infty$ above, the wave function is a non-normalizable eigenstate of $P$, but not $D$.  Clearly no pure state can be found that is an eigenstate of both $P$ and $D$.  By taking $a<\infty$, the above state is normalizable and physical, with controllably small fluctuations in the value of $P$.  We don't think the situation improves much for mixed states: the only possibility invariant under both $P$ and $D$ is the ``identity matrix" in the center of mass coordinate, but it is dubious whether (even representing just one coordinate in the $N\rightarrow\infty$ limit) such a highly non-normalizable state could be considered.  This paragraph simply re-states, in a concrete model, what we already noted  -- it is not possible to simultaneously diagonalize $D$ and $P$.  

Is this a physical effect?  Since the center of mass coordinates \emph{completely decouple} from $H$, they can arguably be removed from Hilbert space entirely without loss of generality.  What remains in the Hilbert space are the long-lived fluctuations of $\pi^i(k)$, which are subject to the same Mermin-Wagner fluctuations mentioned above.  There is therefore no notion of long range order measurable by correlations of local operators that are not group-invariant: in $d=1$, $\langle (p_1-p_{N/2})^2 \rangle \sim L$.  

To throw one more wrench into the mix, however, there is a critical difference between SSB of compact U(1) and non-compact dipole symmetry (alone, isomorphic to $\mathbb{R}$).  In the U(1) case, the Goldstone $\varphi$ is not singly-valued: one must look at the well-posed operators $A_m = \mathrm{e}^{\mathrm{i}m\varphi}$ for $m\ne 0$.  These operators have $\langle A_m\rangle = 0$ implying no SSB for $d\le 2$ in any physical ensemble.  However for dipole symmetry, the global momentum $P$ and its average $\pi^i(k=0) = \frac{1}{N}P$ are perfectly well-defined operators. We have constructed normalizable (ground!) states in which the physical operator $\langle \pi\rangle$ takes on a finite value, which shifts under dipole transformations.  In this sense, the large fluctuations assured by the Mermin-Wagner Theorem seem less dangerous as in a conventional U(1) superfluid, and in fact the Mermin-Wagner theorem is argued to only apply to compact symmetry groups \cite{dobrushin_1975}.   However, like in the superfluid, there can be no long-range coherence in $\pi$, which exhibits large local fluctuations.  We leave open the question of whether this is a semantic or a crucial physical difference.

\section{Hydrodynamics with energy conservation}\label{sec:energy}

We now discuss how to incorporate energy conservation into our hydrodynamic theory.  We will need another hydrodynamic degree of freedom $X^0(\sigma^t,\sigma^I)$ that non-linearly realizes the time translation $P^0$, where now $\sigma^t$ is the proper time in the fluid's local rest frame.
This leads to an additional invariant building-block in the fluid spacetime defined as
\begin{align}
    e^0_{s,A}(\sigma) = \frac{\p X^\mu_{s}(\sigma)}{\p \sigma^A}e^0_{s,\mu}(\sigma)
\end{align}
In order to distinguish from solid phases, the effective theory must be invariant under time-independent reparametrizations of the proper time $\sigma^t$,
\begin{align}
    \sigma^t\to \sigma^t+f(\sigma^I),
\end{align}
which implies that the invariant $r$-field without derivatives can only be $e^0_t$. We then introduce the \emph{proper} temperature
\begin{align}
    \beta\equiv e^0_t,
\end{align}
as the thermodynamic temperature\footnote{All the $\beta_0$ appeared in the previous sections must be taken to be $\beta(\sigma)$ as a function of spacetime. We automatically take that into account in the following derivation without lengthy repetitions.}. In the classical limit and physical spacetime, the corresponding $a$-field can be written as
\begin{align}
    e^0_{a,A} = \p_A X^\mu E^0_{a,\mu},\quad E^0_{a,\mu} = e^0_{a,\mu}+\CL_{X_a}e^0_\mu.
\end{align}
Then, we can include a term $T^\mu_0 E^0_{a,\mu}$ into the leading effective Lagrangian \eqnref{eq:linearL}, and by variation with respect to $X_a^0$, we get a new Ward identity, i.e. the energy conservation equation,
\begin{align}
    \p_\mu T^\mu_0 +A^b_0 e_{\mu b} J^\mu - F^b_{0\mu}J^\mu_b - F_{0\mu }J^\mu=0.
\end{align}
The dynamical KMS transformation for the energy variables is given by
\begin{subequations}
\begin{align}
    \tilde{e}^0_\mu(-x)&= e^0_\mu(x),\\
    \tilde{E}^0_{a,\mu}(-x) & = E^0_{a,\mu}(x)+\ii \CL_\beta e^0_\mu (x),
\end{align}
\end{subequations}
with $\CL_{\beta}e^0_\mu = \p_\mu \beta $.

We now generalize the fluid Lagrangian to include energy conservation, focusing on quadratic order in perturbations (i.e., linear response), and switching off background fields. Keeping into account dynamical KMS symmetry, the new terms in the total Lagrangian that depend on $E_{a,\mu}^0$ are $\CL_{\varepsilon} = \CL_{\mathrm{eq},\varepsilon}+\CL_{\varepsilon}^{(1)}+\CL_{\varepsilon}^{(2)}$, with 
\begin{subequations}\label{eq:Lwenergy}
\begin{align}
    \CL_{\mathrm{eq},\varepsilon} &= -(\varepsilon+p) u^\mu E^0_{a,\mu} + p e^\mu_0 E^0_{a,\mu} , \\ 
    -\ii \beta \CL_{\varepsilon}^{(2)} &= \kappa^{ij}E^0_{a,i}E^0_{a,j}+2\alpha^{ij}C_{a,i} E_{a,j}^0,\\
    \beta \CL_{\varepsilon}^{(1)} & = -\kappa^{ij}\CL_{\beta}e^0_i E^0_{a,j}-\alpha^{ij}( C_{a,i}\CL_{\beta}e^0_j+\CL_{\beta}B_i E^0_{a,j}),\label{eq:lep1}
\end{align}
\end{subequations}
where $\kappa^{ij} = \kappa\delta^{ij}$ is the isotropic thermal conductivity, $\alpha^{ij}=\alpha\delta^{ij}$, $\alpha,\kappa\geq 0$, and where we defined the energy density through $\varepsilon+p = -\beta \frac{\p P}{\p \beta}$. In the above, we are taking $\vep=\vep_0+\delta\vep$ and $p=p_0+\chi_\vep^{-1}(\vep_0+p_0)\delta\vep+\chi^{-1}n_0\delta n$, where $\chi_\vep=-\beta\frac{\p\vep}{\p\beta}$ is the specific heat, and $\vep_0,p_0$ are background values of energy density and pressure. The other possible $O(a^2)$ contributions in $\CL_{\varepsilon}^{(2)}$ come with a time derivative, so we can eliminate them through a frame redefinition as explained below (\ref{eq:additionL}). 
At linear order, the ideal charge and dipole currents in (\ref{eq:idealTJ}) are not modified, and thus we can still use eq. (\ref{eq:uvarphi}) to eliminate $u^i$ in favor of the dipole Goldstone $\varphi_b$. On the other hand, eq. (\ref{eq:intXi}) is updated to
\be 0=\p_0\left[C_{a,i}+\frac{\hat{\rho}}{n_0}E^i_{a,0}-\frac{\vep_0+p_0}{n_0}E^0_{a,i}\right]\ee
where the last term above comes from the first term in $\mathcal L_{\text{eq,}\vep}$. Eliminating $C_{a,i}$ will produce terms proportional either to $E_{a,0}^i$, or to $E^0_{a,i}$. The former contributions can be eliminated through a frame redefinition, following similar steps to the discussion below (\ref{eq:intXi}). The latter will renormalize the value of the thermal conductivity $\kappa$ (preserving dynamical KMS invariance) and can also be eliminated.\footnote{To see this, note that $\mathcal L_\beta B_i$, appearing in the last term in (\ref{eq:lep1}), can be replaced using $n_0\mathcal L_\beta B_i=(\vep_0+p_0)\mathcal L_\beta e_i^0$, which can be inferred from the ideal part of the momentum conservation equation.} We can thus set $\alpha=0$.
By varying with respect to the $e^0_{a,\mu}$, we find the equilibrium energy-4-current as 
\begin{align}
    T^\mu_{(0)0} = -\varepsilon u^\mu - p(u^\mu - e^\mu_0)\,. 
\end{align}
Unlike the momentum density, the energy density is still governed by the fluid elements
; on the algebraic level, this is because the time translation and the dipole symmetry commute. The dissipative part is the usual temperature gradient:
\begin{align}
    T^i_{(1)0} = -\kappa^{ij}\beta^{-1}\p_j \beta\,,
\end{align}
and the noise contribution to the energy current is $\tau^i$, with $\langle\tau^{i}(x)\tau^{j}(0) \rangle = 2T_0 \kappa^{ij} \delta^{(d+1)}(x)$.

We now analyze the normal modes. 
Keeping into account corrections to \eqnref{eq:nonlTJ}, $
     T^{ib}_{(0),\varepsilon}\approx \chi_\vep^{-1}(\vep_0+p_0)\delta \varepsilon
    \,\delta^{ib}
$, and substituting the expressions found above into the Ward identities, we obtain equation of motions for the hydrodynamic modes and the dipole Goldstone. The complete expression of the normal modes is complicated due to the additional coupling to the energy sector. Thus, we do not present the full solutions here for the sake of conciseness but emphasize on few consequences after including the energy sector. First, the $d-1$ subdiffusive modes in \eqnref{eq:subdiff} continue to exist but with a different subdiffusion constant. Second, the energy diffusion will generically mix with the magnon-like sound mode since they both scale as $\sim k^2$. To see it, we keep the dissipative hydrodynamics in the energy sector but ideal hydrodynamics in the momentum and charge sectors. The resulting equation of motion is given by, in the linear response,
\begin{subequations}\label{eq:eomen}
\begin{align}
    \p_0 \delta \varepsilon  + n_0^{-1}(\varepsilon_0+p_0)a^{ijkc}\p_i\p_j\p_k\varphi_c -\chi_\varepsilon^{-1} \kappa^{ij}\p_i\p_j \delta \varepsilon&=0\\
    n_0\p_0 \varphi^b+\chi^{-1}n_0 \p_i \delta n \delta^{ib}+\chi_\varepsilon^{-1}(\varepsilon_0+p_0) \p_i \delta \varepsilon \delta^{ib}&=0 \\
    \p_0 \delta n-a^{ijkb}\p_i\p_j\p_k \varphi_b &=0\,.
\end{align}
\end{subequations}
After doing a Fourier transform, the normal modes are the solutions of the following cubic equation
\begin{align}
    \omega^3+\ii\frac{\kappa}{\chi_{\varepsilon}}\omega^2k^2+ \left(\frac{a}{\chi_\varepsilon} \left(\frac{\varepsilon_0+p_0}{n_0}\right)^2 - \frac{a}{\chi}\right) \omega k^4 -\ii \frac{a\kappa}{\chi\chi_\varepsilon}k^6 = 0
\end{align}
The resulting modes are 3-fold: two propagating modes $\omega=\pm ck^2 - \ii \gamma k^2$ and one diffusion mode $\omega = -\ii \gamma^{\prime}k^2$, where explicit expressions for $c,\gamma,\gamma^\prime$ are not illuminating. The propagating modes still have the magnon-like dispersion but the leading dissipative contribution is now $\sim-\ii k^2$. The diffusion mode is reminiscent of the energy diffusion in a normal fluid.

We conclude by noting that, because of the diffusive nature of the three longitudinal modes above, it is not clear a priori whether the hydrodynamic instability and the associated non-Gaussian universality class emergent at long wavelengths found in \cite{Glorioso:2021bif} will survive at long times, or whether it will be replaced by a different universality class. This is an interesting question that we leave for future work.


\section{From Galilean symmetry to dipole symmetry}\label{sec:gal}
Lastly, let us point out an interesting connection between the physics described above, and the $m\rightarrow\infty$ (infinite mass) limit of the Galilean symmetry algebra.  This was described in a different formalism in the recent work \cite{Kapustin:2022fzp}.  Consider a system preserving both time $H$ and space $P_i$ translations, $\mathrm{U}(1)$ charge $Q$, and Galilean boost $K_i$.  Similar to the dipole symmetry, Galilean boosts are also broken spontaneously in hydrodynamics \cite{nicolisSSB,komargodski_boost} due to the following algebra:
\begin{subequations}\label{eq:boost}
\begin{align}
    [K_i,P_j]&= \ii m Q\delta_{ij},\\
    [K_i,H]&= \ii P_i,
\end{align}
\end{subequations}
where $m$ is the mass of underlying particles.  Due to the second commutator above, we note that the resulting algebra is distinct from the dipole-momentum algebra we discuss in this paper.  Still, observe that in conventional liquids and gases there are no obvious ``Goldstone bosons": the hydrodynamics are described by simply charge, energy and momentum conservation.\footnote{In the literature it is remarked that the Goldstone of broken boosts is the conventional sound wave \cite{komargodski_boost}, but we remark that even without any boost or rotational symmetries, such sound waves still exist \cite{Huang:2022ixj}.  So the sound wave is not crucially reliant on the broken continuous boost symmetry, while the ``sound mode" of the dipole-momentum theory cannot be disentangled from symmetry breaking, as far as we can tell.}

Following general constructions in \appref{app:curved}, we obtain the invariant blocks in the flat spacetime as 
\begin{subequations}
\begin{align}
    e^\mu_A &= \p_A X^\mu -\p_A X^0 \delta^\mu_i \eta^i , \\
    V^i_A &= \p_A \eta^i ,\\
    B_A &=\p_A \varphi + m \p_A X^i  \eta_i-\frac{1}{2}m \p_A X^0 \eta^i\eta_i ,
\end{align}
\end{subequations}
where we associated the Goldstone $\eta^i$ to the Galilean boost symmetry $K_i$.
Following the construction described in Section \ref{sec:EFT}, the $r$-fields invariant under relabeling symmetries are $e^\mu_t$, $B_t$ and $V^i_\mu$ (upon index contraction). We can also derive the $a$-fields in the classical limit. For our purpose, we will ignore nonlinear terms associated with $X^\mu_a$ but keep relevant terms for $\eta^i_{a}$. The resulting $a$-fields are
\begin{subequations}
\begin{align}
    e^\mu_{a,A}&= \p_A X^\nu E^\mu_{a,\nu},\quad E^\mu_{a,\nu} \approx \p_\nu X^\mu_a-\delta^0_\nu \delta^\mu_i \eta^i_a,\\
    V^i_{a,A} & =\p_A X^\mu V^i_{a,\mu} ,\quad V^i_{a,\mu}= \p_\mu \eta^i_a,\\
    B_{a,A}&=  \p_A X^\mu C_{a,\mu},\quad C_{a,\mu}\approx \p_\mu \varphi_a+m\delta^i_\mu \eta_{a,i}-m\delta^0_\mu \eta^i \eta_{a,i},
\end{align}
\end{subequations}
To the leading order in $a$-fields, the effective Lagrangian is given by
\begin{align}
    \CL = \hat{T}^\mu_\nu E^\nu_{a,\mu} +J^\mu C_{a,\mu}+W^{\mu}_i V^i_{a,\mu}+\ldots.
\end{align}
Varying with respect to $\eta_{a,i}$, we obtain the boost Ward identity 
\begin{align}\label{eq:boostWI}
     -\hat{T}^0_i+mJ_i-m J^0 \eta_i - \p_\mu W^\mu_i=0.
\end{align}
Writing the stress tensor and currents in terms of thermodynamic variables, $\hat{T}^0_i\sim u_i+\eta_i$, $J_i\sim n_0 (u_i+\eta_i)$, $J^0\sim n_0$ and $\p_\mu W^\mu_i\sim \CO(\p^2\eta_i)$, we find that 
\begin{align}\label{eq:etau}
    \eta_i \sim u_i+\CO(\p^2).
\end{align}
This immediately tells that the boost Goldstones are \emph{redundant} degree of freedoms.
More explicitly, if including the first-order dissipative effect
\begin{align}\label{etadamp}
    \beta \CL^{(1)}\sim -\Gamma \CL_\beta e^i_{0} E^i_{a,0}\supset -\Gamma\beta \p_0\eta^i\eta_{a,i},
\end{align}
we see that the boost equation of motion \eqnref{eq:boostWI} will be damped at the timescale $\sim \Gamma$.
Therefore, the broken boost does not provide additional hydrodynamic modes, and the long-time dynamics is equivalent to an ordinary (boost-invariant) charge fluid. As a bonus, by setting $\eta_i=0$ (as it will decay to this value), \eqnref{eq:boostWI} implies that the momentum density is equal to the mass current: 
\begin{align}
    T^0_i|_{\eta_i=0}\equiv \left(\hat{T}^0_i + mJ^0\eta_i\right)|_{\eta_i=0}=  mJ_i|_{\eta_i=0}.
\end{align}

Note however that, as in the dipole fluid, the Galilean boost with parameter $c_i$ causes a shift to the momentum density: \begin{equation}
    T^0_i \rightarrow T^0_i + m n_0 c_i .
\end{equation}
This is well known from textbook fluid mechanics -- it is simply the statement that hydrodynamics is the same in all inertial reference frames, and that momentum transforms from one such frame to another.  The fact that $[K_i,P_i]\sim Q$ is mathematically analogous to $[D_i,P_i]\propto Q$ explains why the dipole shift symmetry causes such a similar transformation.  Unlike the dipole symmetry however, $[K_i,H]\ne 0$ and this causes the $\eta_i$ ``dipole Goldstone" to \emph{also} show up in $E^\mu_{a,0}$ as well as its quadratic form in $C_{a,0}$, which causes $\eta_i$ to relax.

Remarkably, there exists an well-defined limit for the Galilean boost symmetry -- taking the infinite mass limit $m\to \infty$ and keeping $D_i\equiv K_i/m$ (but not $K_i$ itself) a good symmetry generator. To keep everything in the same order, the boost Goldstone needs to be scaled as $\varphi_i\equiv m\eta_i$ since $\eta_i\sim m^{-1}\to 0$. In this limit, the boost algebra reduces exactly to the dipole algebra by identifying $D_i$ and $\varphi_i$ as the dipole generator and Goldstone correspondingly. Moreover, by defining $\p_\mu J^\mu_i\equiv \p_\mu W^\mu_i/m$ as the divergence of dipole current and upon charge conjugation, the boost Ward identity \eqnref{eq:boostWI} becomes the dipole Ward identity \eqnref{eq:WIdipolemain}.
Consequently, the relation \eqnref{eq:etau} is incomplete at the leading order since $\eta_i\to 0$, and going to the next leading order, we find it reproduces the dipole constraint \eqnref{eq:uvarphi}. So long as $\Gamma$ remains finite, the dissipation term in \eqnref{etadamp} vanishes since $\eta_a \rightarrow m^{-1}\varphi_a \rightarrow 0$.   The (dipole) Goldstone can no longer be integrated out; in fact, as seen before, it is the velocity which becomes redundant\footnote{The number of degrees of freedom is nevertheless the same in both cases, and it is smaller than that to start with -- we are losing $d$ redundant modes in spatial dimension $d$. The reduction of the true degrees of freedom is a common feature (not yet proven) for spacetime symmetry breaking based upon the inverse Higgs mechanism \cite{ivanov_1975_inverse,Low:2001bw,McArthur:2010zm,Nicolis:2013sga}, therefore,
it is interesting to understand to what extend can we generalize this statement.}. This reveals how the algebraic observation that the Galilean algebra becomes the multipole algebra at $m=\infty$ is realized in the EFT.   



We are not sure whether or not existing methods \cite{Son:2005rv,Son:2008ye,Son:2013rqa,Jensen:2014aia,jensen_galilean} used to study ``non-relativistic conformal field theories" (with Galilean symmetry) can be neatly used in the infinite mass limit; this could be a fruitful direction for future research.

The physics discussed above might be useful to study physics in strongly correlated flat bands, in the regime where the infinite mass limit is exact \cite{Vishwanath_flat_2019prl}.

\section{Conclusions}
In this paper, we have developed an effective field theory for fluids with dipole and momentum conservation.  Our construction highlights a few subtle aspects of this problem: in particular, it appears necessary to write down a local action in terms of more degrees of freedom than are actually present in the effective theory, despite the absence of an obvious need for Lagrange multipliers.  A crucial observation that arises out of this construction is that the dipole symmetry is generally spontaneously broken, \emph{and} that the Goldstone boson for this broken symmetry is essentially the momentum density.

The fact that dipole symmetry is spontaneously broken is also found in \cite{Jensen:2022iww}.  Unlike in that reference however, we find this conclusion is deeply related to having momentum conservation; without momentum, there is no need to have spontaneously broken the dipole symmetry.  That this effect appears common in the large $N$ models of \cite{Jensen:2022iww} may be an artifact of the solution method in the large $N$ limit.  At the very least, there is no evidence for spontaneous symmetry breaking of dipole symmetry in any classical (Markov chain) models of dipole-conserving hydrodynamics studied thus far; in contrast, the numerical simulations of \cite{Glorioso:2021bif} did find compelling evidence for the universality class whose field theory was derived in the present paper.   It would be interesting to understand this issue further in future work.

An important lesson from this effective field theory construction was the generalization from flat to curved background.  Here, our approach perhaps differs conceptually from other attempts in the recent literature \cite{jensencurved}.  In our approach, we followed recent work \cite{Huang:2022ixj} which emphasized the importance of using vielbein indices (not spatial indices) to encode conservation laws in curved space: this construction made it possible to couple anisotropic fluids to geometry.  Since in principle it may be desirable to study anisotropic dipole- and momentum-conserving fluids, we expect that such a vielbein construction is also preferable here.  More importantly, by starting with a first-order formulation, it was natural to assert that in curved space one should relax the requirement that the gauge field is a mixed rank object ($A_t,A_{ij}$), and to instead simply require a pair of gauge fields $A_\mu$ and $A_\mu^b$ corresponding to charge and dipole. The mixed-rank gauge field can then only emerge in a flat space limit. Moreover, the vielbein formalism helps us to understand two key physical consequences: the existence of the dipole Goldstones and the asymmetric part of the dipole fields.

Looking forward, we anticipate that our methods can be generalized to discover infinitely many new hydrodynamic universality classes that arise in fracton-like classical or quantum matter.  It may be straightforward conceptually, if tedious in practice, to generalize this construction to include higher multipole conservation laws.  A more important and interesting direction will be to understand how to generalize the geometrically inspired construction presented here to non-thermal matter -- after all, the highlight of this work is the dynamics without energy conservation!  Recent work \cite{Guo:2022ixk} along these lines has begun, but the consequences or diagnosis of spontaneous symmetry breaking in this new approach have not yet been understood. 

It will also be fascinating to look for experimental realizations of the dipole and momentum conserving hydrodynamics developed here, whether in high quality solid-state devices in very large electric fields \cite{sachdev_2002}, or in low density interacting ultracold atoms in tilted trapped optical lattices \cite{guardado2020subdiffusion}.  We also hope that progress along these lines will be made in the next few years.  It may also be possible to explore similar (though it appears distinct) fixed points which arise from the symmetry group of volume-preserving diffeomorphisms (which can arise in lowest Landau level physics)  \cite{Du:2021hes,Du:2022xys}; this algebra is equivalent to the dipole-momentum algebra at the linearized level.

\section*{Acknowledgements} 
We acknowledge useful discussions with Anton Kapustin, Rahul Nandkishore, Shu-Heng Shao, Dam Thanh Son, Lev Spodyneiko, and especially Kristan Jensen and Nathan Seiberg.  PG and AL thank the Simons Center for Geometry and Physics for hospitality.

This work was supported by the Department of Energy through Award DE-SC0019380 (PG), the Simons Foundation through Award No. 620869 (PG), the National Science Foundation under CAREER Award DMR-2145544 (XH, JG, AL), the Gordon and Betty Moore Foundation's EPiQS Initiative via Grants GBMF4302 and GBMF8686 (JFRN), and GBMF10279 (XH, JG, AL), and by Research Fellowships from the Alfred P. Sloan Foundation under Grant FG-2020-13615 (PG) and FG-2020-13795 (AL).

\begin{appendix}

\section{Memory matrix methods} \label{sec:MM}
In this appendix, we use the memory matrix formalism \cite{forster2018hydrodynamic} to derive the linearized hydrodynamics of the main text, both near and away from charge neutrality.  This approach provides an independent check on many of the non-trivial properties of hydrodynamics that we found above and can give some interesting perspectives on the results.

The memory matrix formalism is an old set of formal manipulations, used to isolate the contributions to linear response theory (two-point functions) which arise from parametrically slow dynamics.  Since long wavelength hydrodynamic modes are arbitrarily long lived, this method can be well-suited for calculations of their properties.  We now tersely summarize the main results of this method: for details see \cite{forster2018hydrodynamic}. Consider a many-body system with Hamiltonian $H$, at temperature $T$.  One can construct a vector space consisting of all  operators $A$ acting on this system: to emphasize the vector nature, we can write $|A)$.   An inner product on this space is \begin{equation}
    (A|B) := T\int\limits_0^\beta \mathrm{d}\lambda \langle A^\dagger(\mathrm{i}\lambda)B\rangle_T
\end{equation}
with $T=1/\beta$ and $\langle \cdots \rangle_T = \frac{1}{\mathrm{tr}(\mathrm{e}^{-\beta H})} \mathrm{tr}(\mathrm{e}^{-\beta H}\cdots )$ the thermal expectation value.   Note that the susceptibility matrix is \begin{equation}
    (A|B) = T\chi_{AB}.
\end{equation}

Suppose that we have a designated set of ``slow" operators $|\mathcal{O}_A)$.  For us, these are naturally taken to be $n(\mathbf{k})$ and $\pi^i(\mathbf{k})$ (the Fourier wave number is $\mathbf{k}$, and is held fixed).  We may define the projectors \begin{equation}
    \mathfrak{p} = \sum_{\text{slow} A,B} |A) (T\chi)^{-1}_{AB} (B|, \;\;\;\;\; \mathfrak{q}=1-\mathfrak{p},
\end{equation}
which project degrees of freedom onto slow ($\mathfrak{p}$) and fast ($\mathfrak{q}$) modes.  By noticing that $(A| (z-\mathcal{L})^{-1}|B)$ is linearly related to the retarded Green's function $G^{\mathrm{R}}_{AB}(z)$, one can show that there are hydrodynamic quasinormal modes whenever \cite{} 
\begin{equation}
    \det (M + N - \mathrm{i}\omega \chi ) = 0.
\end{equation}
Here $M$ (the memory matrix) and $N$ are given by \begin{subequations}
\begin{align}
    M_{AB} &=  (\dot{A}| \mathfrak{q} \mathrm{i}(z-\mathfrak{q}\mathcal{L}\mathfrak{q})^{-1}\mathfrak{q}|\dot{B}), \\
    N_{AB} &= -N_{BA} = \chi_{\dot{A}B}.
\end{align}
\end{subequations}
Here $\mathcal{L} = \mathrm{i}[H,\cdot]$ denotes the Liouvillian, and $\dot{A} = \mathrm{i}[H,A]$, with $H$ the overall Hamiltonian.  

In this paper, we aim to use this framework to gain further insight (and justification) for the non-trivial hydrodynamics discovered in Section \ref{sec:EFT}.  Strictly speaking, one can object to this on the grounds that energy conservation is explicit in any theory satisfying the above postulates.  Ultimately, we will use this approach to discern what happens when energy is conserved along with dipole and momentum; however, we believe that this approach remains instructive even if we ``ignore" energy conservation as an unjustified assumption.  As we will see, some of the confusing features of this fluid are consequences of very general, and even semi-microscopic, arguments.

\subsection{Momentum susceptibility}
Let us begin by determining the momentum susceptibility; in the memory matrix language, this is $(\pi|\pi) = T \chi_{\pi \pi}$ (we'll leave the Fourier index implicit for the remainder of this section).  While a microscopic computation is not possible (nor important for hydrodynamic considerations), we can easily \emph{bound} susceptibility using the Cauchy-Schwarz inequality: \begin{equation}
    (\pi_x|\pi_x) \ge \frac{(\pi_x|J_x)^2}{(J_x|J_x)}.
\end{equation}
Here $J_x$ is the $x$-component of the charge current operator; for simplicity, we'll also take $\mathbf{k}=k\hat{\mathbf{x}}$.  Now, observe two key properties.  Firstly, in a generic many-body system, \begin{equation}
    (\pi_x|J_x) = T n_0, 
\end{equation}
with $n_0$ the equilibrium charge density: $n_0=\langle n\rangle_T$. A heuristic and fully general argument for this fact, which we have not seen in the literature, is as follows: \begin{equation}
    \frac{(\pi_x|J_x)_{k\rightarrow 0}}{T} = \int_0^\infty \mathrm{d}t \int \frac{\mathrm{d}^dx}{V} \; \mathrm{i}\langle [J_x(x,t),P_x]\rangle = \int_0^\infty \mathrm{d}t \int \frac{\mathrm{d}^dx}{V} \; \langle \partial_x J_x\rangle = -\int_0^\infty \mathrm{d}t \int \frac{\mathrm{d}^dx}{V} \; \langle \partial_t n\rangle = Tn_0
\end{equation}where in the last step we have used integration by parts.  This argument is not rigorous because if the $k\rightarrow 0$ limit is taken too quickly the integral trivially vanishes.  Secondly, using \eqnref{eq:WIdipolemain}, \begin{equation}
    (J_x|J_x) = k^2 (J_{xx}|J_{xx}).
\end{equation}
Since $J_{xx}$ is the local current operator which is well-defined with local dipole conservation, we conclude that  $(J_{xx}|J_{xx}) $ is $k$-independent as $k\rightarrow 0$, and should remain finite as $n_0\rightarrow 0$.  Combining these 3 equations, we find that for some constant $c>0$, which does not vanish as $n_0\rightarrow 0$, \begin{equation}
    \chi_{\pi\pi} = c \frac{n_0^2}{k^2},
\end{equation}
in agreement with the result \eqnref{eq:momsusc} of the main text.

\subsection{Dynamics without energy}
Now, let us consider the dynamics without energy conservation.   In this case, we'll include $\pi_i$ and $n$ as degrees of freedom in the memory matrix formalism.   The non-zero matrix elements of the $N$ matrix are \begin{equation}
    N_{\pi_i n} = N_{n\pi_i} = (\pi_i | \dot{n}) = \mathrm{i}k_j (\pi_i | J_j) = \mathrm{i}k_j \delta_{ij} Tn_0 = T \times \mathrm{i}k_i n_0.
\end{equation}
The most important non-zero matrix elements of the $M$ matrix are \begin{subequations}\begin{align}
    M_{nn} &= (\dot{n}|\mathfrak{q} \mathrm{i} (\omega - \mathfrak{q}\mathcal{L}\mathfrak{q})^{-1}\mathfrak{q}|\dot{n}) = k_i k_j k_k k_l (J_{ij}| \mathrm{i} (\omega - \mathfrak{q}\mathcal{L}\mathfrak{q})^{-1}|J_{kl}) = T\alpha k^4, \\
    M_{\pi_i \pi_j} &= (\dot{\pi}_i|\mathfrak{q} \mathrm{i} (\omega - \mathfrak{q}\mathcal{L}\mathfrak{q})^{-1}|\dot{\pi}_j) = k_k k_l (T_{ik}| \mathfrak{q} |T_{jl}) = T\left(\beta k_i k_j + \gamma k^2 \delta_{ij}\right).
\end{align}\end{subequations}
for some constants $\alpha,\beta,\gamma$.   The hydrodynamic normal modes thus come from solving \begin{equation}
    0 = \det(M+N-\mathrm{i}\omega \chi ),
\end{equation}
which in the transverse sector gives 
\begin{subequations}
\begin{align}
    \omega \chi_{\pi\pi} &= \gamma k^2,\\
    \omega &= -\mathrm{i}\frac{\gamma k^4}{cn_0^2},
\end{align}
\end{subequations}
and in the longitudinal sector gives \begin{equation}
    \det \left(\begin{array}{cc} \alpha k^4 -\mathrm{i}\omega \chi_{nn} &\ \mathrm{i}kn_0 \\ \mathrm{i}kn_0 &\ (\beta+\gamma)k^2 -\frac{\mathrm{i}\omega cn_0^2}{k^2} - \mathrm{i}\omega A \end{array} \right)
\end{equation}
Here $A>0$ is a finite constant that persists even as $n_0\rightarrow 0$, arising from $\chi_{\pi\pi}$   Indeed, if $n_0=0$, then we see that both longitudinal and transverse momentum get $k^2$ decay rates, and charge has $k^4$ subdiffusion, while if $n_0>0$, we have \begin{equation}
    0 = k^4 n_0^2 + (\alpha k^4 - \mathrm{i}\omega \chi_{nn}) ((\beta+\gamma)k^4 - \mathrm{i}\omega cn_0^2)
\end{equation}
which is solved by \begin{equation}
    \omega = \pm \frac{k^2}{\sqrt{c\chi_{nn}}} - \mathrm{i}\Gamma k^4 + \cdots
\end{equation}
in agreement with the prediction \eqnref{eq:normalmom} of the main text.

\section{Dipole fluids with momentum in a curved spacetime}\label{app:curved}

In this section, we will discuss the change of dipole hydrodynamics by coupling to a curved spacetime. Such analysis also tells us how to source various currents in the flat spacetime limit as discussed in \secref{sec:EFT}. It is interesting to notice that because of the non-commutativity between the dipole moment and the momentum operator, the momentum density is not invariant under dipole shifts. However, since the energy operator commutes with the dipole moment, the energy density is invariant. Consequently, our theory forbids any type of boost symmetries between space and time, and a natural choice of spacetime to describe such theory is the Aristotelian background \cite{Boer2018,Jain2021,jensencurved,Bradlyntopo}. 
It is most natural to use vielbein formalism to describe the geometry, thus we introduce the internal (flat) spacetime indices $\alpha,\beta$ and $b,c$ for its spatial subspace (we reserve $a$ to describe $a$-fields in the Keldysh contour!). The dipole and spacetime algebra is given by
\begin{subequations}\label{eq:algebracurve}
\begin{align}
    [P_b,D_c]&= -\ii Q\delta_{bc},\label{eq:PDQcurve}\\
    [P_d,L_{bc}] &= \ii (\delta_{dc}P_b - \delta_{db}P_c),\\
    [D_d,L_{bc}] &= \ii (\delta_{dc}D_b - \delta_{db}D_c),\\
    [L_{bc},L_{de}] & = \ii (\delta_{bd}L_{ce} -\delta_{be}L_{cd} - \delta_{cd}L_{be} +\delta_{ce}L_{bd}),
\end{align}
\end{subequations}
where $L_{bc}$ generates the $SO(d)$ rotational symmetry. The time translation $P_0$ commutes with every other generator.

Before constructing the field theory, it is instructive to define the dipole moment operationally as 
\begin{align}\label{eq:defDcurve}
    D_b \equiv  \int \mathrm{d}^d x e~ y_b(x^i) n(x),
\end{align}
for some arbitrary function $y_b(x^i)$ in terms of spatial coordinates.  In particular, let us assume for now that the coordinates $\lbrace y^b\rbrace$ form the ``natural" coordinates where the vielbein would be constant: $e^\mu_b = \delta^\mu_b$.  Nevertheless, we will not for now invoke any such constraint.   We also emphasize that on a general curved background, $y_b$ is not globally well-defined, with the 2-sphere the simplest example.   This means that one should only understand the resulting theory as a `covariant' way of coupling a dipole and momentum conserving theory to curved space -- the explicit coupling to the metric ends up breaking the conservation laws.  This phenomenon is not unusual, and is well-known to happen already with momentum conservation on a generic curved background.  In what follows, we will treat derivatives on the vielbein at the same order as derivatives on the hydrodynamic fields; thus e.g. the curvature scalar $R\sim \mathcal{O}(k^2)$.

The time derivative of the dipole moment implies 
\begin{align}
    \p_0 D_b = \int \mathrm{d}^d x e~ y_b(x^i) \p_0n(x) = -  \int \mathrm{d}^d x ~ y_b(x^i)  \p_k(e J^k) =  \int \mathrm{d}^d x e~  \p_k (y_b(x^i))J^k\triangleq \int \mathrm{d}^d x e~ e_{kb}J^k,  \label{eq:p0Db}
\end{align}
where we have used the covariant charge conservation in \eqnref{eq:WIQcurve}. The last equation defines the vielbein $e_{k}^b\equiv \p_k y^b$ that give us the transformation from $y^b$ to the uglier coordinates $x^i$, such that the dipole moment will be conserved according to the dipole Ward identity in \eqnref{eq:WIdipolecharge}. 
We emphasize that the existence of such a choice of viebein is not generic: in particular, the Ricci curvature tensor vanishes. Therefore, only on a flat space can the dipole moment be defined in terms of operators in \eqnref{eq:defDcurve}. 

To build the invariant blocks and to include the spontaneous dipole symmetry breaking, we apply the \emph{coset construction} for the spacetime symmetries \cite{Landry2020, delacrtaz2014} (for an introduction to the coset construction, see the references therein).
Let us focus on dipole fluids without energy conservation. The unbroken generators are the translations $P_\alpha$, and the charge $Q$; the broken generator is the dipole moment $D_b$. Thus, the most general group element is parametrized as
\begin{equation}
    g(\sigma) =e^{\ii \beta_0\sigma^t P_0} e^{\ii y^b(X(\sigma))P_b} e^{\ii \varphi^b(\sigma)D_b}e^{- \frac{\ii}{2}\theta^{b c}(\sigma)L_{b c}}e^{\ii \varphi(\sigma)Q},
\end{equation}
where $\sigma^A$ represent the fluid spacetime, and we associate with each symmetry generator a dynamical field. $P_0$ is an effective time-translation supporting a fixed temperature $T_0=\beta_0^{-1}$ determined by the noise.
We introduce the gauge fields for translations ($\hat{e}^\alpha_\mu$), rotations ($\omega^b_{\mu c}$), charges ($\hat{A}_\mu$), and dipole moments ($A^b_\mu$) , thus the gauge invariant Maurer-Cartan one-form is given by 
\begin{equation}
    g^{-1}\left(\p_A+\ii\p_A X^\mu\hat{e}^\alpha_{\mu}P_\alpha+ \frac{\ii}{2}\p_A X^\mu \omega^{bc}_{\mu} L_{ b c}+\ii \p_A X^\mu \hat{A}_\mu Q+\ii \p_A X^\mu A^b_\mu D_b \right) g =\ii e_A^\alpha P_\alpha+\ii K^b_{A} D_b  +\ii B_A Q+\ii \frac{1}{2} \Theta^{b c}_{A} L_{ b c}.
\end{equation}
Hence, the useful invariant blocks are given by 
\begin{subequations}
\begin{align}
    e^b_A &= \p_A X^\mu e^c_\mu R^{\ph b}_{c}, \\
    B_A &= \p_A \varphi+\p_A X^\mu A_\mu  + \p_A X^\mu e^b_\mu  \varphi_b , \\
    K^b_A &=  \left(\p_A \varphi^c +\p_A X^\mu \omega^{c}_{\mu d} \varphi^d +\p_A X^\mu A^c_\mu \right)R^{\ph b}_{ c} ,\label{eq:KK}\\
    \Theta^b_{A c} &= \p_A X^\mu \left[(R^{-1})^{bd}\p_\mu R_{d c} - (R^{-1})^{bd}\omega^{e}_{\mu d} R_{e c} \right],
\end{align}
\end{subequations}
where to linear order in $\theta^{bc}$ the rotation matrix reads $R^{bc} = \delta^{bc} - \theta^{b c}$, and we defined 
\begin{align}
    e^b_\mu&\equiv \p_\mu y^b(X)+\hat{e}^b_\mu+\omega^{b}_{\mu c} y^c(X),\\
    A_\mu &\equiv \hat{A}_\mu - A^b_\mu y_b(X).\label{eq:redefA}
\end{align}
Clearly, we should regard $e^b_\mu$ as the vielbein, $\omega^{b}_{\mu c} = -\omega^{c}_{\mu b}$ as the spin connection, and $A_\mu$ as the $\mathrm{U}(1)$ gauge field. Together with $e^0_\mu$ (which can be chosen arbitrarily, e.g. simply set to $\delta^0_\mu$),\footnote{Note however that this vielbein does not act as the ``gauge field" coupled to energy current, in contrast to $e^b_\mu$ and the momentum current.} the vielbein $e^\alpha_\mu$ is invertible and satisfies the orthogonality and completeness relations
\begin{align}
    e_\alpha^\mu e_\mu^\beta=\delta_\alpha^\beta, \quad e_\alpha^\mu e_\nu^\alpha=\delta_\nu^\mu.
\end{align}
The vielbein formalism requires also the Christoffel connection $\Gamma^\rho_{\mu\nu}$, and the covariant derivative is defined as 
\begin{subequations}
\begin{align}\label{eq:c9a}
    \nabla_\mu e^0_\nu &= \p_\mu e^0_\nu - \Gamma^{\rho}_{\mu\nu} e^0_\rho,\\
    \nabla_\mu e^b_\nu &= \p_\mu e^b_\nu +\omega^b_{\mu c}e^c_\nu - \Gamma^{\rho}_{\mu\nu} e^b_\rho.
\end{align}
\end{subequations}
The spin connection and Christoffel connection are not independent, so we impose the metric compatibility
\begin{align}
    \nabla_\mu e^\alpha_\nu = 0,
\end{align}
and treat the vielbeins and the spin connections as the independent background fields with respect to which the action would vary.
We note a useful relation 
\begin{align}\label{eq:dett}
    \Gamma^\mu_{\nu\mu} = e^{-1}\p_\nu e,
\end{align}
where $e\equiv \det e^\alpha_\mu$.

To incorporate the blocks into the Schwinger-Keldysh formalism, we introduce the two-time copies ($s=1,2$):
\begin{subequations}\label{eq:curvedblock}
\begin{align}
    e^b_{s,A}(\sigma) &= \frac{\p X^\mu_s(\sigma)}{\p \sigma^A} e^c_{s,\mu}(\sigma) R^{\ph \ph \ph b}_{s,c}(\sigma), \\
    B_{s,A}(\sigma) &= \frac{\p X^\mu_s(\sigma)}{\p \sigma^A}\left( A_{s,\mu}(\sigma)+e^b_{s,\mu}(\sigma)\varphi_{s,b}(\sigma) \right)+\frac{\p \varphi_s(\sigma)}{\p \sigma^A},\\
    K_{s,A}^b(\sigma)&=  \frac{\p X^\mu_{s}(\sigma)}{\p \sigma^A}\left(A_{s,\mu}^c(\sigma)+\omega_{s,\mu d}^c(\sigma)\varphi_s^d(\sigma) \right)R^{\ph \ph \ph b}_{s,c}(\sigma)+\frac{\p \varphi^c_s(\sigma)}{\p \sigma^A}R^{\ph \ph \ph b}_{s,c}(\sigma),\\
    \Theta^b_{s,A c}(\sigma) &=  \frac{\p X^\mu_{s}(\sigma)}{\p \sigma^A} \left[(R_s^{-1})^{bd}\p_{s,\mu} R_{s,d c} - (R_s^{-1})^{bd}\omega^{e}_{s,\mu d} R_{s,e c} \right](\sigma),
\end{align}
\end{subequations}
Restricting to $\theta^{bc}_r=0$, the $r$-fields are 
\begin{subequations}\label{eq:curvedmomblock}
\begin{align}
    e^b_{r,A} &= \p_A X^\mu e^b_\mu , \\
    B_{r,A} &= \p_A \varphi+\p_A X^\mu A_\mu  + \p_A X^\mu e^b_\mu  \varphi_b , \\
    K^b_{r,A} &=  \p_A \varphi^b +\p_A X^\mu \omega^{b}_{\mu d} \varphi^d +\p_A X^\mu A^b_\mu  ,\label{eq:KK} \\
    \Theta^b_{r,Ac} & = \p_A X^\mu \omega^b_{\mu c}.
\end{align}
\end{subequations}
In the classical limit and physical spacetime, the $a$-fields are given by
\begin{subequations}
\begin{alignat}{3}
    e^b_{a,A}&= \p_A X^\mu E^b_{a,\mu},\quad  && E^b_{a,\mu}= e^b_{a,\mu}+\CL_{X_a}e^b_\mu+e^c_\mu \theta^b_{a,c}, \\
    B_{a,A}&= \p_A X^\mu C_{a,\mu},   &&C_{a,\mu} = A_{a,\mu}+\p_\mu \varphi_a+\CL_{X_a}A_\mu+e^b_\mu \varphi_{a,b}+(e^b_{a,\mu}+\CL_{X_a}e^b_\mu)\varphi_b, \\
    K^b_{a,A} &= \p_A X^\mu K^b_{a,\mu},  && K^b_{a,\mu}=\nabla_\mu \varphi^b_a +\omega^b_{a,\mu c}\varphi^c+\varphi^c\CL_{X_a}\omega^b_{\mu c}  + A^b_{a,\mu}+\CL_{X_a}A^b_\mu+K^c_\mu \theta^b_{a,c},\\
    \Theta^b_{a,A c}&= \p_A X^\mu \Omega^b_{a,\mu c},\quad  && \Omega^b_{a,\mu c} = -\nabla_\mu \theta^b_{a,c} +\omega^b_{a,\mu c}+\CL_{X_a}\omega^b_{\mu c}\,,
\end{alignat}
\end{subequations}
where $\nabla_\mu \theta_{a,c}^b=\p_\mu\theta_{a,c}^b+\omega^b_{\mu d}\theta_{a,c}^d-\omega^d_{\mu c}\theta_{a,d}^b$.
It is straightforward to check that the invariant blocks defined above reflect the consistency of symmetry algebra in \appref{app:consistency}. 
To the leading order in $a$-fields, the effective Lagrangian can be written as
\begin{align}\label{eq:effLmom}
    \CL = \hat{T}^\mu_b E^b_{a,\mu} +J^\mu C_{a,\mu}+J^{\mu}_b K^b_{a,\mu}+ \hat{S}^{\mu c}_{ \ph \ph \ph b} \Omega^b_{a,\mu c}+\ldots.
\end{align}
%
%
%
%
To derive the momentum Ward identity, let us consider the variation with respect to $X^\nu_a$. Denoting compactly the total stress tensor $T^\mu_b \equiv \hat{T}^\mu_b+J^\mu\varphi_b$ and the total spin current $S^{\mu c}_{\ph \ph \ph b}\equiv \hat{S}^{\mu c}_{\ph \ph \ph b}+J^\mu_{[b} \varphi^{c]}$, we obtain
\begin{align}
    \frac{\delta S}{\delta X_a^\nu} & =-e^{-1}\p_\mu [e (T^\mu_b e^b_\nu +S^{\mu c}_{\ph \ph \ph b}\omega^b_{\nu c}+J^\mu_b A^b_\nu + J^\mu A_\nu)] +T^\mu_b \p_\nu e^b_\mu+ S^{\mu c}_{\ph \ph \ph b}\p_\nu \omega^b_{\mu c} + J^\mu_b \p_\nu A^b_\mu  +J^\mu\p_\nu A_\mu  \nonumber\\
    & = -e^{-1}\p_\mu(e T^\mu_b)e^b_\nu  -e^{-1}\p_\mu(e S^{\mu c}_{\ph \ph \ph b})\omega^b_{\nu c}  - e^{-1}\p_\mu(e J^\mu_b)A^b_\nu - e^{-1}\p_\mu(e J^\mu)A_\nu \nonumber\\
    &\quad + 2T^\mu_b \p_{[\nu}e^b_{\mu]}+2S^{\mu c}_{\ph \ph \ph b} \p_{[\nu}\omega^b_{\mu] c}  + 2J^\mu_b \p_{[\nu} A^b_{\mu]}+2J^\mu \p_{[\nu}A_{\mu]}\notag \\
    & = -\nabla'_\mu(T^\mu_b) e^b_\nu -\nabla'_\mu (J^\mu_b) A^b_\nu -\nabla'_\mu J^\mu A_\nu+T^\mu_b G^b_{\nu\mu}+S^{\mu c}_{\ph \ph \ph b} R^b_{\ph c \nu\mu}+J^\mu_b F^b_{\nu\mu}+J^\mu F_{\nu\mu},\label{eq:covmomEOM}
\end{align}
where we defined the modified covariant derivative $\nabla'_\mu\equiv \nabla_\mu+G^\rho_{\mu\rho}$ with $G^\lambda_{\mu\nu} \equiv 2\Gamma^\lambda_{[\mu\nu]}$
, and the field strength
\begin{subequations}\begin{align}
    F_{\mu\nu} &\equiv \p_\mu A_\nu - \p_\nu A_\mu,\\
    F^b_{\mu\nu} &\equiv \p_\mu A^b_\nu - \p_\nu A^b_\mu +\omega^b_{\mu c}A^c_\nu - \omega^b_{\nu c}A^c_\mu,\\
    G^b_{\mu\nu} &\equiv \p_\mu e^b_\nu - \p_\nu e^b_\mu+\omega^b_{\mu c} e^c_\nu - \omega^b_{\nu c} e^c_\mu,\\
    R^b_{\ph c\mu\nu} &\equiv \p_\mu \omega^b_{\nu c} - \p_\nu \omega^b_{\mu c}+\omega^b_{\mu d} \omega^d_{\nu c} - \omega^b_{\nu d}\omega^d_{\mu c}\label{eq:riem}.
\end{align}\end{subequations}
In the last step of \eqnref{eq:covmomEOM}, we used the spin current Ward identity obtained by varying the action with respect to $\theta^{bd}_a$:
\begin{align}
    T^\mu_{[b} e^{d]}_\mu + J^\mu_{[b} A^{d]}_\mu + \nabla'_\mu S^{\mu d}_{\ph \ph \ph b}=0\,, \label{eq:rotEOM}
\end{align}
as well as the dipole Ward identity \eqnref{eq:WIdipolecharge}.
It is known, however, that the ``intrinsic'' spin current $\hat{S}^{\mu c}_{\ph \ph \ph b}$ is a non-hydrodynamic mode \cite{Glorioso_nonAbelian}. This can be seen through $-\p_0 \hat{S}^{0 c}_{\ph \ph \ph b}\sim  \hat{S}^{0 c}_{\ph \ph \ph b} \subset \hat{T}^\mu_{[b} e^{c]}_\mu$ which means that $\hat{S}^{0 c}_{\ph \ph \ph b}$ will relax to zero at long time.\footnote{$\hat{S}^{0 c}_{\ph \ph \ph b}$ is identified as the local angular momentum density in \cite{Glorioso_nonAbelian}.} Since the spatial part $\hat{S}^{i c}_{\ph \ph \ph b}$ is proportional to (gradient of) $\hat{S}^{0 c}_{\ph \ph \ph b}$, we are allowed to set $\hat{S}^{\mu c}_{\ph \ph \ph b}=0$ in the following.\footnote{Note that in an ordinary fluid without dipole symmetry, imposing $\hat{S}^{\mu c}_{\ph \ph \ph b}=0$ to \eqnref{eq:rotEOM} would lead to a symmetric stress tensor $\hat{T}^\mu_{[b}e^{d]}_\mu=0$.} On the other hand, the ``non-intrinsic'' spin current $S^{\mu c}_{\ph \ph \ph b}=J^\mu_{[b}\varphi^{d]}$ is not relaxed.  
Last, the variation with respect to $\varphi_a$ and $\varphi^b_a$ will give charge and dipole Ward identities. Combining them, we obtain, with ordinary derivatives,
\begin{subequations}
\begin{align}
    e^{-1}\p_\mu \left( e (\hat{T}^\mu_b+J^\mu\varphi_b)\right)+ e^{-1}\p_\mu \left( e J^\mu_d\varphi^c\right)\omega^d_{\nu c}e^\nu_b +e^{-1}\p_\mu(e J^\mu_c)A^c_\nu e^\nu_b  &\nonumber\\
    - 2(\hat{T}^\mu_c+J^\mu\varphi_c) \p_{[\nu}e^c_{\mu]}e^\nu_b - 2 J^\mu_d\varphi^c \p_{[\nu}\omega^d_{\mu]c}e^\nu_b - 2 J^\mu \p_{[\nu}A_{\mu]}e^\nu_b -2J^\mu_c \p_{[\nu} A^c_{\mu]} e^\nu_b &=0,\\
    e^{-1}\p_\mu(e J^\mu)
    &=0,\label{eq:WIQcurve}\\
    e^{-1}\p_\mu (e  J^{\mu}_b) - \omega^c_{\mu b}J^\mu_c  -J^\mu e_{\mu b}  
    &=0,
\end{align}
\end{subequations}
and with covariant derivatives, 
\begin{subequations}\label{eq:WIcurve}
\begin{align}
    \nabla'_{\mu}\left(\hat{T}^\mu_b+J^\mu\varphi_b\right)+A^c_\nu e_{\mu c}J^\mu e^\nu_b- R^d_{\ph c \nu \mu}\varphi^c J^\mu_d e^\nu_b - F^d_{\nu \mu}J^\mu_d e^\nu_b - F_{\nu\mu}J^\mu e^\nu_b - G^c_{\nu\mu} \left(\hat{T}^\mu_c+J^\mu\varphi_c\right) e^\nu_b &=0,\label{eq:WImomcurve} \\
    \nabla'_\mu J^\mu
    &=0,\label{eq:WIQcurve}\\
    \nabla'_\mu  J^{\mu}_b  -J^\mu e_{\mu b}  
    &=0.\label{eq:WIdipolecharge}
\end{align}
\end{subequations}
These are the momentum, charge and dipole equations of motions in an generic curved spacetime. Remarkably, the dipole Goldstone will couple to the curvature to give a force in the momentum equation, and we have showed that it is best understood as the Joule heating due to spin currents \cite{Bradlyntopo}.

If we set $\varphi^b=0$, our result is consistent with \cite{jensencurved} in that the structure of their Eq.(4.13) is recovered: $\nabla'_\mu \hat{T}^\mu_b = f_\mu e^\mu_b - e^0_\mu \hat{T}^\mu_c e^c_\rho e^\nu_b\nabla_\nu e^\rho_0 $, where we denoted collectively the Joule heating term $f_\mu$ and used $G^b_{\mu\nu} = 2 e^b_\lambda e^0_{[\mu}\nabla_{\nu]}e^\lambda_0$.
However, unlike \cite{jensencurved}, our formalism includes the dipole Goldstone $\varphi^b$, which leads to the coupling between spin current and background curvature. Note that this feature is ultimately related to the breakdown of dipole gauge theory in curved spacetime \cite{slaglecurve, gromovcurve}, but here the dipole symmetry is valid so long as we count background curvature as derivative corrections.
There is another conceptual difference from \cite{jensencurved} that our $A^b_\mu$ needs not to be symmetric. This is possible if we include the dipole Goldstone $\varphi^b$. Unlike \cite{jensencurved}, the antisymmetric part of $A^b_\mu$ is not fixed by $F_{\mu\nu}$, and does have physical consequences as emphasized in the main text.

\section{Consistency of the symmetry algebra}\label{app:consistency}
The analysis of this section follows largely \cite{jensencurved} but contains several generalizations of it. 

Let $\chi^\mu$ be an infinitesimal diffeomorphism, $\Omega^b_{\ph c}$ an infinitesimal rotation, $\Lambda$ a $U(1)$ gauge transformation, and $\xi^b$ a dipole shift parameter. 
We start by defining the action of transformations $\Xi=(\chi^\mu,\Omega^b_{\ph c},\Lambda,\xi^b)$ on the dynamical fields $X^\mu(\sigma),\varphi(\sigma),\varphi^b(\sigma)$. We have
\be e^{\delta_\Xi}X^\mu =(1+\delta_\Xi+\cdots)X^\mu=X^\mu - \chi^\mu(X)+\cdots\,,\ee
thus, in leading order,
\be [\delta_{\Xi'},\delta_\Xi]=[e^{\delta_{\Xi'}},e^{\delta_\Xi}]\,.\ee
We have, up to second order,
\be e^{\delta_{\Xi'}}e^{\delta_\Xi} X^\mu = X^\mu - \chi^\mu-\chi'^\mu+\chi'^\nu\p_\nu\chi^\mu,\ee
so that\footnote{We take passive transformations on dynamical fields, i.e. $[\delta_{\Xi'},\delta_{\Xi}] = -\delta_{[\Xi',\Xi]}$.}
\be [e^{\delta_{\Xi'}},e^{\delta_\Xi}]X^\mu = \mathcal L_{\chi'}\chi^\mu = -  \delta_{[\Xi',\Xi]}X^\mu,\ee
where we carefully kept into account zeroth, first, and second-order terms and verified that zeroth and first-order terms cancel out. We then find
\be \chi^\mu_{[\Xi',\Xi]}\equiv \delta_{\Xi'}\chi^\mu=\mathcal L_{\chi'}\chi^\mu\,.\ee
Next, let's look at $\varphi^b$. We have
\be e^{\delta_\Xi}\varphi^b=\varphi^b+\xi^b - \Omega^b_{\ph c}\varphi^c.\ee
Note that we are \emph{not} including $\mathcal L_\chi \varphi^b$ in the above as we view $\varphi^b$ as a function of $\sigma$, which is a singlet under physical spacetime diffeomorphisms $\chi^\mu$. Then, 
\be e^{\delta_{\Xi'}}e^{\delta_\Xi} \varphi^b=\varphi^b+\xi^b+\xi'^b - (\Omega'^b_{\ph c}+\Omega^b_{\ph c})\varphi^c-\Omega^b_{\ph c}\xi'^c+\Omega^b_{\ph d}\Omega'^d_{\ph\ph c}\varphi^c-\CL_{\chi'} \xi^b+\CL_{\chi'}\Omega^b_{\ph c}\varphi^c,\ee
and
\be [e^{\delta_{\Xi'}},e^{\delta_\Xi}]\varphi^b=-\CL_{\chi'}\xi^b + \CL_{\chi}\xi'^b + \Omega'^b_{\ph c}\xi^c - \Omega^b_{\ph c}\xi'^c+ (\CL_{\chi'}\Omega^b_{\ph c} - \CL_{\chi}\Omega'^b_{\ph c} )\varphi^c +(\Omega^b_{\ph d}\Omega'^d_{\ph c} -\Omega'^b_{\ph d}\Omega^d_{\ph c} )\varphi^c, \ee
thus giving
\begin{subequations}\begin{align}
    \xi^b_{[\Xi',\Xi]}&\equiv\delta_{\Xi'}\xi^b=\CL_{\chi'}\xi^b - \CL_{\chi}\xi'^b + \Omega^b_{\ph c}\xi'^c - \Omega'^b_{\ph c}\xi^c\,,\\
    \Omega^b_{\ph c,[\Xi',\Xi]}&\equiv\delta_{\Xi'}\Omega^b_{\ph c}= \CL_{\chi'}\Omega^b_{\ph c} - \CL_{\chi}\Omega'^b_{\ph c}  +\Omega^b_{\ph d}\Omega'^d_{\ph c} -\Omega'^b_{\ph d}\Omega^d_{\ph c} .
\end{align} \end{subequations}
Now it's $\varphi$'s turn:
\be\label{eq:varyvarphi}
e^{\delta_\Xi}\varphi=\varphi+\Lambda\equiv\varphi+\lambda+M^b \xi_b + N_\mu \chi^\mu\,,\ee
where $M^b$ and $N_\mu$ are expressions in terms of $X^\mu,\varphi$ and $\varphi^b$. Up to a normalization of the transformation parameters, the most general choice consistent with charge conjugation invariance is 
\be
M^b= (1-c) e^b_\mu X^\mu,\quad N_\mu =c e_\mu^b \varphi_b\,.
\ee
We will determine $c=0$ later from consistency requirements, but now it can be any value. Again we view $\varphi$ as a function of $\sigma$ and thus we do not have the term $\mathcal L_\chi \varphi$. Composing two transformations,
\be e^{\delta_{\Xi'}}e^{\delta_\Xi}\varphi=\varphi+\Lambda+\Lambda'-\CL_{\chi'}\Lambda + c e^b_\mu \chi^\mu \xi'_b
\,,\ee
gives
\begin{align}
    [e^{\delta_{\Xi'}},e^{\delta_\Xi}]\varphi& = -\CL_{\chi'}\Lambda + \CL_{\chi}\Lambda'+c e^b_\mu \chi^\mu \xi'_b-c e^b_\mu \chi'^\mu \xi_b  
\end{align}
To summarize this part, the algebra of local transformations is
\begin{subequations}\label{eq:deltaXi}
\begin{align} \chi^\mu_{[\Xi',\Xi]}\equiv& \delta_{\Xi'}\chi^\mu=\mathcal L_{\chi'}\chi^\mu,\\
\Omega^b_{c,[\Xi',\Xi]}\equiv &\delta_{\Xi'}\Omega^b_{\ph c}= \CL_{\chi'}\Omega^b_{\ph c} - \CL_{\chi}\Omega'^b_{\ph c}  +\Omega^b_{\ph d}\Omega'^d_{\ph c} -\Omega'^b_{\ph d}\Omega^d_{\ph c} ,\\
\xi^b_{[\Xi',\Xi]}\equiv&\delta_{\Xi'}\xi^b=\CL_{\chi'}\xi^b - \CL_{\chi}\xi'^b + \Omega^b_{\ph c}\xi'^c - \Omega'^b_{\ph c}\xi^c, \\
\Lambda_{[\Xi',\Xi]}\equiv&\delta_{\Xi'}\Lambda
= \CL_{\chi'}\Lambda - \CL_{\chi}\Lambda'-c e^b_\mu \chi^\mu \xi'_b+c e^b_\mu \chi'^\mu \xi_b .
\end{align}
\end{subequations}
This is a generalization of Eq.(4.11) in \cite{jensencurved}.
To see the consistency with the dipole algebra, we decompose the field variation into
\begin{align}
    \delta_\Xi = - \ii \chi^\mu P_\mu + \ii \lambda Q + \ii \xi^b D_b,
\end{align}
where $P_\mu$, $Q$ and $D_b$ are symmetry generators, and we have ignored the rotation symmetry generator for simplicity. The background field is also turned off. As a consequence of \eqnref{eq:varyvarphi}, we have 
\begin{align}
    \ii Q \varphi = 1,\quad \ii D_b\varphi = M_b,\quad -\ii P_\mu \varphi = N_\mu.
\end{align}
Now, let us take $\Xi' = \xi'_b$ and $\Xi = \chi^\mu$, then
\begin{align}
    -[\delta_{\Xi'},\delta_{\Xi}]\varphi = -\chi^\mu \delta^b_\mu \xi'_c [D_b,P_c]\varphi - (1-c) X^\mu \delta^b_\mu \chi^\rho\p_\rho \xi'_b.
\end{align}
At the same time, we have
\begin{align}
    -[\delta_{\Xi'},\delta_{\Xi}]\varphi = \Lambda_{[\xi'_b,\chi^\mu]} = -\chi^\mu \delta^b_\mu \xi'_b- (1-c) X^\mu \delta^b_\mu \chi^\rho\p_\rho \xi'_b.
\end{align}
Thus we conclude that
\begin{align}
    [P_b,D_c] = -\ii Q\delta_{bc}
\end{align}
is the desired dipole algebra.

There is an unwanted free parameter $c$ appearing in the algebra. Here, we show that in order for the transformation $\Xi$ itself to also form a closed algebra, one must set $c=0$. First, as a warm-up, let us consider the variation of $\chi^\mu,\xi^b,\Omega^b_{\ph c}$. We have
\begin{align}
    [\delta_{\Xi''},\delta_{\Xi'}]\chi^\mu = \CL_{\chi''}\CL_{\chi'}\chi^\mu - \CL_{\chi'}\CL_{\chi''}\chi^\mu = \CL_{\chi_{[\Xi'',\Xi']}}\chi^\mu = \delta_{[\Xi'',\Xi']} \chi^\mu.
\end{align}
Next, using product rules, e.g. $\delta_{\Xi''}\CL_{\chi'}\xi^b =\CL_{\delta_{\Xi''}\chi'}\xi^b+ \CL_{\chi'}\delta_{\Xi''}\xi^b $ , we find
\begin{align}
    \delta_{\Xi''}\delta_{\Xi'}\xi^b = & \Omega''^b_{\ph\ph c}\CL_{\chi}\xi'^c+\Omega^b_{\ph c}\CL_{\chi''}\xi'^c -\CL_{\chi''}\Omega'^b_{\ph c}\xi^c - \CL_{\chi}\Omega'^b_{\ph c}\xi''^c + \Omega^b_{\ph c} \Omega'^c_{\ph d} \xi''^d +\Omega''^b_{\ph\ph d}\Omega'^d_{\ph c}\xi^c \nonumber\\
    &+\left(\CL_{\chi''}\Omega^b_{\ph c}\xi'^c + \CL_{\chi'}\Omega^b_{\ph c}\xi''^c - \Omega''^b_{\ph\ph c}\CL_{\chi'}\xi^c -\Omega'^b_{\ph c}\CL_{\chi''}\xi^c -\Omega''^b_{\ph\ph d} \Omega^d_{\ph c}\xi'^c - \Omega'^b_{\ph d} \Omega^d_{\ph c}\xi''^c  \right) , 
\end{align}
thus
\begin{align}
    [\delta_{\Xi''},\delta_{\Xi'}]\xi^b = \CL_{\chi_{[\Xi'',\Xi']}}\xi^b - \CL_{\chi}\xi^b_{[\Xi'',\Xi']} +  \Omega^b_{\ph c} \xi^c_{[\Xi'',\Xi']}  - \Omega^b_{\ph c,[\Xi'',\Xi']} \xi^c  = \delta_{[\Xi'',\Xi']} \xi^b.
\end{align}
Similar calculation leads to
\begin{align}
    [\delta_{\Xi''},\delta_{\Xi'}]\Omega^b_{\ph c} = \CL_{\chi_{[\Xi'',\Xi']}}\Omega^b_{\ph c} - \CL_{\chi}\Omega^b_{\ph c,[\Xi'',\Xi']} +  \Omega^b_{\ph d} \Omega^d_{\ph c,[\Xi'',\Xi']} - \Omega^b_{\ph d,[\Xi'',\Xi']} \Omega^d_{\ph c}  = \delta_{[\Xi'',\Xi']} \Omega^b_{\ph c}.
\end{align}
Now, we look at $\Lambda$. For brevity, let us take $e^b_\mu=\delta^b_\mu$ and temporarily turn off $\Omega^b_{\ph c}$. Note, this is merely a simplification for manipulations and the final result should still hold in arbitrarily curved spacetime. We have
\begin{align}
    \delta_{\Xi''}\delta_{\Xi'}\Lambda &= \delta_{\Xi''}(\CL_{\chi'}\Lambda - \CL_{\chi}\Lambda'-c \delta^b_\mu \chi^\mu \xi'_b+c \delta^b_\mu \chi'^\mu \xi_b) \notag\\
    & = \CL_{\CL_{\chi''}\chi'}\Lambda+\CL_{\chi'}(\CL_{\chi''}\Lambda - \CL_{\chi}\Lambda'' - c\delta^b_\mu (\chi^\mu \xi''_b- \chi''^\mu \xi_b)) - \CL_{\CL_{\chi''}\chi}\Lambda' - \CL_{\chi}(\CL_{\chi''}\Lambda' - \CL_{\chi'}\Lambda''- c\delta^b_\mu (\chi'^\mu \xi''_b- \chi''^\mu \xi'_b))\nonumber\\
    &\;\;\;\; - c\delta^b_\mu \CL_{\chi''}\chi^\mu \xi'_b  - c\delta^b_\mu \chi^\mu (\CL_{\chi''}\xi'_b - \CL_{\chi'}\xi''_b) + c\delta^b_\mu \CL_{\chi''}\chi'^\mu \xi_b  + c\delta^b_\mu \chi'^\mu (\CL_{\chi''}\xi_b - \CL_{\chi}\xi''_b) \nonumber\\
    & = \CL_{\chi''}\CL_{\chi'}\Lambda +\CL_{\chi}\CL_{\chi'}\Lambda'' -(\CL_{\chi''}\CL_{\chi}\Lambda' +\CL_{\chi'}\CL_{\chi}\Lambda'') +c\delta^b_\mu \left[\CL_{\chi}\chi'^\mu \xi''_b - \CL_{\chi}(\chi''^\mu \xi'_b) - \chi^\mu \CL_{\chi''}\xi'_b \right] \nonumber \\
    &\;\;\;\;+c\delta^b_\mu \left[   - (\CL_{\chi'} \chi^\mu \xi''_b+ \CL_{\chi''} \chi^\mu \xi'_b)+ (\CL_{\chi'} \chi''^\mu \xi_b+\CL_{\chi''} \chi'^\mu \xi_b ) + (\chi''^\mu\CL_{\chi'}  \xi_b+\chi'^\mu\CL_{\chi''}  \xi_b)\right],
\end{align}
thus
\begin{align}
    [\delta_{\Xi''},\delta_{\Xi'}]\Lambda &= \CL_{\chi_{[\Xi'',\Xi']}}\Lambda - \CL_{\chi}(\CL_{\chi''}\Lambda' -\CL_{\chi'}\Lambda'')\nonumber\\
    &\;\;\;\; +c\delta^b_\mu\left[ -\CL_{\chi}(\chi''^\mu \xi'_b - \chi'^\mu \xi''_b) - \chi^\mu(\CL_{\chi''}\xi'_b-\CL_{\chi'}\xi''_b) +\CL_{\chi}\chi'^\mu \xi''_b - \CL_{\chi}\chi''^\mu \xi'_b\right].
\end{align}
We note that the r.h.s. cannot be written as $\CL_{\chi_{[\Xi'',\Xi']}}\Lambda -\CL_{\chi}\Lambda_{[\Xi'',\Xi']} - c\delta_{\mu b} \chi^\mu \xi^b_{[\Xi'',\Xi']}+c\delta_{\mu b} \chi^\mu_{[\Xi'',\Xi']} \xi^b$, hence, the algebra is not closed. However, if we set $c=0$, we have
\begin{align}
    [\delta_{\Xi''},\delta_{\Xi'}]\Lambda = \CL_{\chi_{[\Xi'',\Xi']}}\Lambda - \CL_{\chi}\Lambda_{[\Xi'',\Xi']} = \delta_{[\Xi'',\Xi']} \Lambda.
\end{align}
To summarize, when $c=0$, $\Xi$ itself forms a closed algebra
\begin{align}
    [\delta_{\Xi''},\delta_{\Xi'}]\Xi = \delta_{[\Xi'',\Xi']} \Xi.
\end{align}

Lastly, let us turn to the background gauge fields. The variation is defined as 
\begin{subequations}
\begin{align}
    \delta_{\Xi} e^b_\mu & = \CL_{\chi} e^b_\mu-\Omega^b_{\ph c} e^c_\mu,\\
    \delta_{\Xi} \omega^b_{\mu c}&=\CL_{\chi}\omega^b_{\mu c}+ \nabla_\mu \Omega^b_{\ph c},\\
    \delta_{\Xi} A_\mu & = \CL_{\chi} A_\mu - \p_\mu \Lambda - e^b_\mu \xi_b,\\
    \delta_{\Xi} A^b_\mu &= \CL_{\chi}A^b_\mu - \nabla_\mu \xi^b - \Omega^b_{\ph c} A^c_\mu.
\end{align}
\end{subequations}
Composing two transformations, we have
\begin{subequations}
\begin{align}
    \delta_{\Xi'}\delta_{\Xi} e^b_\mu &= \CL_{\chi'}\CL_{\chi}e^b_\mu + (\CL_{\chi}\Omega'^b_{\ph c} + \Omega'^b_{\ph d}\Omega^d_{\ph c})e^c_\mu - \left(\CL_{\chi}(\Omega'^b_{\ph c} e^c_\mu) + \CL_{\chi'}(\Omega^b_{\ph c} e^c_\mu)\right),\\
    \delta_{\Xi'}\delta_{\Xi} A_\mu & = \CL_{\chi'}\CL_{\chi}A_\mu + \p_\mu \CL_{\chi}\Lambda' + e^b_\mu \CL_{\chi}\xi'_b - e^b_\mu \Omega_{bc}\xi'^c - \left(\CL_{\chi}\p_\mu\Lambda' +\CL_{\chi'}\p_\mu\Lambda  + \CL_{\chi'}(e^b_\mu \xi_b)+ \CL_{\chi}(e^b_\mu \xi'_b)\right),\\
    \delta_{\Xi'}\delta_{\Xi} \omega^b_{\mu c} & =\CL_{\chi'}\CL_{\chi}\omega^b_{\mu c} -\nabla_\mu (\CL_{\chi}\Omega'^b_{\ph c} + \Omega'^b_{\ph d}\Omega^d_{\ph c})+(\CL_{\chi}\nabla_\mu \Omega'^b_{\ph c}+\CL_{\chi'}\nabla_\mu \Omega^b_{\ph c}+\p_\mu \Omega^b_{\ph d}\Omega'^d_{\ph c} + \p_\mu \Omega'^b_{\ph d}\Omega^d_{\ph c}\nonumber\\
    &\quad +\omega^b_{\mu e}\Omega^e_{\ph d}\Omega'^d_{\ph c} + \omega^b_{\mu e}\Omega'^e_{\ph d}\Omega^d_{\ph c} -\Omega'^b_{\ph e}\omega^e_{\mu d}\Omega^d_{\ph c} - \Omega^b_{\ph e}\omega^e_{\mu d}\Omega'^d_{\ph c} ).
\end{align}
\end{subequations}
The variation of $A^b_\mu$ is a bit tedious, but taking lessons from previous manipulations, we can immediately see that the infinitesimal rotation should be closed. Hence, we are allowed to focus on the piece involving $\xi^b$ only. In particular,
\begin{subequations}
\begin{align}
    \delta_{\Xi'}\delta_{\Xi} A^b_\mu & \supset \nabla_\mu \CL_{\chi}\xi'^b + \p_\mu(\Omega'^b_{\ph c} \xi^c) - \omega^b_{\mu d}\Omega^d_{\ph c}\xi'^c -\left(\CL_{\chi}(\nabla_\mu \xi'^b)+\CL_{\chi'}(\nabla_\mu \xi^b) + \p_\mu\Omega^b_{\ph c} \xi'^c +\p_\mu \Omega'^b_{\ph c} \xi^c +\Omega^b_{\ph c}\omega^c_{\mu d}\xi'^d +\Omega'^b_{\ph c}\omega^c_{\mu d}\xi^d  \right).
\end{align}
\end{subequations}
Then, it is straightforward to check that $[\delta_{\Xi'},\delta_{\Xi}]  = \delta_{[\Xi',\Xi]}$ holds for the background fields as well:
\begin{subequations}
\begin{align}
    [\delta_{\Xi'},\delta_{\Xi}] e^b_\mu & = \CL_{\chi_{[\Xi',\Xi]}}e^b_\mu - \Omega^b_{\ph c,[\Xi',\Xi]} e^c_\mu = \delta_{[\Xi',\Xi]} e^b_\mu,\\
    [\delta_{\Xi'},\delta_{\Xi}] A_\mu & =
    \CL_{\chi_{[\Xi',\Xi]}}A_\mu - \p_\mu \Lambda_{[\Xi',\Xi]} - e_{\mu b}\xi^b_{[\Xi',\Xi]}
    = \delta_{[\Xi',\Xi]} A_\mu,\\
    [\delta_{\Xi'},\delta_{\Xi}] A^b_\mu & = \CL_{\chi_{[\Xi',\Xi]}}A^b_\mu -  \nabla_\mu \xi^b_{[\Xi',\Xi]} - \Omega^b_{\ph c,[\Xi',\Xi]}A^c_\mu = \delta_{[\Xi',\Xi]} A^b_\mu,\\
    [\delta_{\Xi'},\delta_{\Xi}] \omega^b_{\mu c} & = \CL_{\chi_{[\Xi',\Xi]}}\omega^b_{\mu c} + \nabla_\mu \Omega^b_{\ph c,[\Xi',\Xi]} = \delta_{[\Xi',\Xi]} \omega^b_{\mu c}.
\end{align}
\end{subequations}

\comment{

\section{Hydrodynamic instabilities} \label{sec:instability}

In this section, we first confirm that the tensors $\zeta,\eta$ defined in \eqnref{eq:tensors} are describing the bulk/shear viscosity in dipole fluids. Then, we show in both the bulk/shear viscosity that the nonlinearity in pressure would lead to instability below $d=4$, which is consistent with the critical dimension discussed at the end of \secref{sec:dissipation}.

The Kubo formula for shear viscosity is\footnote{We shall focus on $d=2$, but in general one can regard $x$ as the longitudinal direction and $y$ as the transverse directions.}
\be \eta_{\mathrm{shear}}=\frac{\beta_0}{2}\lim_{\omega\to 0}\lim_{k\to 0}\langle\{T^{xy}(\omega,k),T^{xy}(-\omega,-k)\}\rangle=\frac{\beta_0}{2}
\lim_{\omega\to 0}\lim_{k\to 0}\left(\frac{\omega}{k_x}\right)^2\langle\{\pi^y(\omega,k_x),\pi^y(-\omega,-k_x)\}\rangle\ ,\ee
where we used that, setting $k_y=0$, $-i\omega \pi^y+ik_x T^{xy}=0$. From the Lagrangian we found, the quadratic part that involves transverse components of momentum $\pi^y=n_0\varphi_y$ is
\begin{align}
    \CL &= n_0 \varphi_i \p_0 X^i_a+\ii \beta_0^{-1}s^{ijkl}\p_i X_{a,j}\p_k X_{a,l} + n_0^{-1} s^{ijkl}a^{mjnp} \p_i\p_m\p_n\varphi_p \p_k X_{a,l} \nonumber\\
    &\to n_0 \varphi_y \p_0 X^y_a+\ii \beta_0^{-1}\eta (\p_x X_{a,y})^2+n_0^{-1}\eta a_2 \p_x^3 \varphi_y \p_x X_{a,y} \nonumber\\
    &\to  \frac 12\begin{pmatrix}\varphi_y^*&X_{a,y}^*\end{pmatrix}
\begin{pmatrix}0 & -i\omega n_0+ n_0^{-1}\eta a_2 k^4_x\\
i\omega n_0+ n_0^{-1}\eta a_2 k^4_x& -2i \beta_0^{-1}\eta k_x^2\end{pmatrix}
\begin{pmatrix}\varphi_y\\X_{a,y}\end{pmatrix}.
\end{align}
We then find
\be\label{eq:phiphi} \langle\{\pi^y(\omega,k_x),\pi^y(-\omega,-k_x)\}\rangle
=2n_0^2\langle\varphi_y(\omega,k_x)\varphi_y(-\omega,-k_x)\rangle
=\frac{2n_0^2 \beta_0^{-1}\eta k_x^2}{|in_0\omega+n_0^{-1}\eta a_2 k_x^4|^2}\ee
which leads to
\be \eta_{\mathrm{shear}}=\eta .\ee
Interestingly, despite the subdiffusive dispersion, the shear viscosity does not vanish. This is because the non-local momentum susceptibility $\rho\sim k^{-2}$ cancels out the additional $k^2$ scaling from subdiffusion. 
We can do a similar discussion for the bulk viscosity. The Kubo formula is
\be \zeta_{\mathrm{bulk}}=\frac{\beta_0}{2}\lim_{\omega\to 0}\lim_{k\to 0}\langle\{T^{xx}(\omega,k),T^{xx}(-\omega,-k)\}\rangle=
\frac{\beta_0}{2}\lim_{\omega\to 0}\lim_{k\to 0}\left(\frac{\omega}{k_x}\right)^2\langle\{\pi^x(\omega,k_x),\pi^x(-\omega,-k_x)\}\rangle\ ,\ee
which is thus nonzero for the same arguments as above, and in particular $\zeta_{\mathrm{bulk}}=\zeta$. 

We now show that both $\zeta_{\mathrm{bulk}}$ and $\eta_{\mathrm{shear}}$ diverge when accounting for fluctuations. Indeed, from \eqnref{eq:nonlTJ}, $T^{xx}$ has a nonlinear contribution $T^{xx}\sim \delta n^2$. Defining $\mathcal T^{xx}(t)=\int \mathrm{d}^d x~ T^{xx}(t,x)$, we can write this in terms of $\delta n$ in momentum space
\be \mathcal T^{xx}=\int \mathrm{d}^dk ~\delta n(k)\delta n(-k)\ .\ee
Now recall that, from charge subdiffusion equation in \eqnref{eq:eommom}, $\langle \delta n(t,k)\delta n(0,-k)\rangle\sim e^{-C t k^4}$ with $C>0$. Hence,
\be \langle \mathcal T^{xx}(t)\mathcal T^{xx}(0)\rangle\sim \int \mathrm{d}^d k~ e^{-2C t k^4}\sim \frac 1{t^{\frac d4}}\ .\ee
Plugging this result in the Kubo formula,
\be \zeta_{\mathrm{bulk}}= \beta_0\lim_{\omega\to 0}\int \mathrm{d}t~ e^{\ii\omega t}\langle \mathcal T^{xx}(t)\mathcal T^{xx}(0)\rangle \sim \lim_{\omega\to 0}\int \mathrm{d}t~ e^{i\omega t}\frac 1{t^{\frac d4}}\sim\lim_{\omega\to 0} \omega^{\frac d4-1}.\ee
$\zeta_{\mathrm{bulk}}$ is then divergent for $d< 4$. Similarly, \eqnref{eq:nonlTJ} gives rise to a nonlinear contribution $T^{xy}\sim k^2\varphi_y(k)\varphi_y(-k)$. Note that unlike $T^{xx}$, which only has nonlinearities in the longitudinal channel, $T^{xy}$ contains both longitudinal and transverse contributions in nonlinearity; here we show it using transverse modes since they are subdiffusive even after including the energy sector. Integrating \eqnref{eq:phiphi}, we find $\langle \varphi_y(t,k)\varphi_y(0,-k)\rangle\sim k^{-2}e^{-C' tk^4}$ with $C'>0$. Following similar steps as above, we arrive at a shear viscosity
\begin{align}
    \eta_{\mathrm{shear}} \sim \lim_{\omega\to 0}\omega^{\frac{d}{4}-1}
\end{align}
that diverges for $d<4$.
}

\end{appendix}

\bibliography{dipolemom}

\end{document}